\documentclass[journal,twoside]{IEEEtran}
\usepackage{authblk}
\usepackage{datetime2}
\usepackage{graphicx}
\usepackage{amsmath}
\usepackage{amssymb}
\usepackage{mathrsfs}
\usepackage{stfloats}
\usepackage{dsfont}
\usepackage{setspace}
\usepackage{epstopdf}
\usepackage{cite}
\usepackage{balance}
\usepackage{xcolor}
\usepackage{booktabs}
\usepackage{pifont}
\usepackage{caption}
\usepackage{subcaption}
\usepackage{bm}
\usepackage{bbm}
\usepackage[linesnumbered, ruled,vlined]{algorithm2e}
\usepackage{algorithmic}
\usepackage{hyperref}  
\usepackage{fancyhdr}

\renewcommand{\QED}{\QEDopen}

\begin{document}
\title{
    Distributed Constrained Resource Management in 6G Networks: A Scalable Hybrid Model-Learning Framework
}

\author{
\authorblockN{Thang X. Vu, \IEEEmembership{Senior Member, IEEE}, Cuong Le, Symeon Chatzinotas, \IEEEmembership{Fellow, IEEE}, and Bj\"orn Ottersten, \IEEEmembership{Fellow, IEEE}}
\thanks{This work is supported by the Luxembourg National Research Fund (FNR), ref. C22/IS/17220888/RUTINE, BRIDGES/2023
/IS/18441334/Pre5GNR and INTER/23/17941203/PASSIONATE. For the purpose of open access, and in fulfilment of the obligations arising from the grant agreement, the authors have applied a Creative Commons Attribution 4.0 (CC BY 4.0) license to any Author Accepted Manuscript version arising from this submission. (\emph{Corresponding author: Thang X. Vu}.)}
\thanks{T. X. Vu, S. Chatzinotas, and B. Ottersten are  with the Interdisciplinary Centre for Security, Reliability and Trust (SnT), University of Luxembourg. Emails: \{thang.vu, symeon.chatsinotas, bjorn.ottersten\}@uni.lu.}
\thanks{C. Le is with the School of Computing, National University of Singapore. Email: cuonglv@comp.nus.edu.sg. This work was performed at SnT.}
}

\markboth{Accepted to IEEE Journal on Selected Areas in Communications}{T. X. Vu \MakeLowercase{\textit{et al.}}: Distributed Constrained Resource Management in 6G Networks}

\maketitle

\begin{abstract}
    Radio Resource Management (RRM) is a fundamental challenge in 6G wireless networks, particularly under dynamic user demands, inter-cell interference, and heterogeneous QoS constraints. Centralized optimization solutions are often infeasible in practice due to the lack of system-wide state information, dynamic conditions, and signaling delays, making distributed learning-based approaches attractive. However, conventional multi-agent reinforcement learning (MARL) struggles with scalability and constraint satisfaction in such highly dynamic environments. We introduce a scalable two-phase hybrid learning framework to address the RRM challenges where orthogonal-frequency division multiplexing (OFDM) domain knowledge is explicitly incorporated into the MARL pipeline. In our proposed two-phase learning framework, the first phase allocates a minimum resource to satisfy users' QoS requirements based on channel statistics, avoiding the inefficiencies of training policies under hard QoS constraints. Subsequently, a multiagent system is employed in the second phase to optimally allocate the remaining resources for system throughput maximization. By exploiting the structural interference information, we propose a scalable MARL algorithm which decomposes the original learning problem into smaller subproblems that can be handled independently, thereby avoiding exponential growth of the action space without compromising performance. Extensive simulations in realistic scenarios with  50MHz bandwidth and different numerologies show that our method significantly outperforms existing optimization and learning baselines, offering up to 40\% throughput improvement, and 100\% constraint satisfaction with minimal resource usage. This work bridges the gap between wireless domain insights and principled MARL techniques, showing that MARL is a powerful tool---but \textit{not all you need}---to address the intricacies of constrained, large-scale distributed RRM in 6G networks. 

\end{abstract}


\section{Introduction} 
\label{sec:intro}
Radio Resource Management (RRM) is a fundamental building block in wireless communication systems, governing how limited spectrum and power resources are allocated to users. RRM in 6G networks is even more essential, and yet challenging due to the multi-layered network topology, dense deployments of base stations (BSs) and heterogeneous quality of service (QoS) requirements. 
Efficient RRM directly affects user experience and overall network capacity. 
%
Classical centralized optimization approaches to RRM \cite{Fatemeh23,shen2018fractional,le2019energy}, while theoretically capable of achieving optimal or near-optimal solutions, suffer from the lack of system-wide state information, prohibitive signaling overhead and latency. 
In practice, no central controller has instant access to all relevant channel and user information. 
For example, optimal RRM in a two-cell cellular system requires two BSs to exchange not only the channel state information (CSI) of their active users, but also CSI of interference channels from one BS to users of the other BS. This is practically impossible because the current 3GPP specifications \cite{3GPP5GNR} do not allow a BS to estimate the CSI to users of other BSs. 
As a result, distributed solutions are not merely desirable for scalability, they are the only viable option. Yet, distributed operation brings its own set of fundamental challenges. Each BS must act based on partial, local observations, with no access to the internal states or operations of other BSs. This lack of coordination, combined with the inherently coupled nature of wireless interference, makes the problem significantly more difficult to manage. In particular, no exact global optimum can be guaranteed based on the limited information available at each BS.

Recently, machine learning (ML) is poised to play a pivotal role as one of the foundational pillars of 6G communication systems. Among ML techniques, Multi-agent reinforcement learning (MARL) has attracted significant attention for distributed RRM thanks to its decentralized decision-making capability and adaptivity to dynamic environment, bypassing the need for explicit network models or centralized control \cite{tang2020deep, zhao2019deep, yang2022distributed, zhang2020deep,  mei2024multi, li2022federated, liang2019spectrum, ji2023multi, naderializadeh2021resource, nasir2019multi}. Despite its promise, in our view, MARL should not and cannot be treated as a black-box tool applied blindly to RRM problems. Instead, it should be carefully designed and tailored to the unique characteristics and constraints of wireless systems. Ignoring the intrinsic structure of the problem, such as interference patterns and stringent QoS requirements, often leads to highly suboptimal or even infeasible outcomes. A principled approach that combines the wireless domain knowledge with MARL is therefore essential for developing practical, scalable RRM strategies in 6G networks.

To illustrate the problem, let us consider the RRM in a downlink system where multiple BSs serve users over a shared frequency bandwidth using Orthogonal-frequency division multiplexing (OFDM). Each BS must assign proper subchannels to its users based solely on local information while satisfying users' QoS constraints. This leads to a very large-scale and distributed combinatorial optimization problem. More importantly, unlike the majority of existing works that assume a fixed user set and static traffic demands, real-world scenarios are inherently stochastic. Users arrive and depart the system unpredictably, while their locations and QoS requirements fluctuate over time. In such practical scenario, the core difficulty lies in simultaneously satisfying strict QoS constraints and maximizing the system performance in a dynamic, distributed, and large-scale system. Standard MARL methods struggle to enforce hard constraints reliably, particularly when user demands and problem sizes evolve over time \cite{Wachi20:ICML:safeRL}. Moreover, even in static scenarios, learning to satisfy constraints while optimizing global utility requires extensive exploration, complex reward shaping and hyperparameter tuning. 
Therefore, a practical solution must explicitly account for such uncertainty and variability in the wireless environment as an intrinsic part of the design.

To address these challenges, we propose a hybrid two-phase model-learning framework that integrates the wireless domain knowledge into the learning process. The key idea behind our proposed hybrid framework is to use a model-based approach in the first phase to tackle the most challenging task in MARL, that is \emph{constraints satisfaction} in dynamic environments. This method is fundamentally different from existing solutions that directly learn to minimize the constraints violation probability through risk-aware learning techniques, such as chance-constraint or CVaR \cite{Vu18RiskSensitive,he2024risk}. We will show later that our proposed risk-handling approach significantly surpasses existing risk-aware learning solutions.  
%
%
In the second phase, the available subchannels are allocated by intelligent agents trained via a powerful cooperative MARL algorithm called QMIX \cite{rashid2020monotonic}.  Unlike existing works employing the QMIX algorithm over the full joint combinatorial action space and resulting in computational infeasibility, we propose a \emph{scalable MARL algorithm} via decomposition.
This decomposition is grounded in the localized structure of interference (\emph{domain knowledge}) in OFDM systems. In particular, since the subchannels within each BS are orthogonal, maximizing the system throughput is equivalent to maximizing the throughput on each subchannel independently assuming users' QoS are satisfied. Our decomposition drastically reduces the sample complexity of the training process. The result is a scalable, constraint-aware framework that adapts effectively to stochastic user dynamics and large-scale deployments.

The main contributions of this paper are summarized as follows:
\begin{itemize}
    \item We consider a large-scale resource allocation problem in which multiple BSs operate over a shared spectrum bandwidth without a centralized coordination. Unlike existing studies \cite{naderializadeh2021resource} that assume static or predictable user demands, we address practical scenarios in which users with stochastic QoS requirements arrive and depart the system unpredictably. This external dynamics, combined with the internal nonstationarity of MARL systems, poses a major challenge to effective and reliable learning policies.
    \item We propose a scalable two-phase learning framework that integrates interference structure of the OFDM transmission into the MARL framework: a model-based greedy allocation phase that deterministically satisfies user QoS constraints, and a cooperative MARL phase that intelligently allocates the remaining resources to maximize the system sum-rate. 
    From the learning methodology, our proposed framework can be viewed similarly to Reinforcement learning with human feedback (RLHF) \cite{kaufmann2024RLHF}: the first phase acts as an expert handling the most challenging task for constraints satisfaction. 
    More importantly, by exploiting the inherent interference structure of OFDM, 
    this design not only reduces sampling complexity of the training, but also yields a reliable and robust solution for the highly dynamic 6G network.
    \item We validate the proposed framework through extensive simulations under realistic scenarios complied with 3GPP standardizations with a full bandwidth of 50MHz and various numerologies \cite{3GPP5GNR}. The results indicate that directly applying constrained MARL leads to poor performance. In contrast, the proposed framework consistently outperforms optimization and learning baselines in terms of system sum-rate, QoS satisfaction, and resource utilization efficiency. In particular, our framework can enhance the system sum-rate by 40\% and achieve 100\% constraint satisfaction while using the least number of active subchannels. 
\end{itemize}

The rest of the paper is organised as follows. Section~\ref{sec:related_work} reviews related works from both optimization and learning methods. Section~\ref{sec:system} describes the system model under consideration and briefly review the conventional MARL solution. Section~\ref{sec:solution} provides the proposed framework in both high-level and detailed design steps. Section~\ref{sec:experiments} presents extensive simulation results and performance assessment. Finally, Section~\ref{sec:conclusion} concludes our contributions and provides discussions. 
\section{Related works} 
\label{sec:related_work}
Before the rise of machine learning, a substantial body of literature existed in which traditional approaches, such as optimization and game theory, were extensively employed to tackle the complex challenges of radio resource management. Optimization techniques, such as convex programming, iterative algorithms based on approximation and relaxation, are frequently used for radio resource allocation \cite{shen2018fractional, zeng2021trajectory, le2019energy, ghanem2019resource, ghanem2020resource, bossy2022energy}. Game theory has become particularly relevant for dynamic and competitive environments, such as spectrum sharing and power control in multi-user networks. Methods like Nash equilibrium and auction-based games are used to model interactions between users or networks, ensuring optimal allocation while minimizing interference \cite{han2005fair, meshkati2009energy, yang2020pricing, nguyen2021hierarchical}. While these methods have proven effective in many scenarios, they are often limited by either assumptions of real-time perfect information, centralization, or the computational complexity of solving large-scale problems. As wireless networks continue to evolve, particularly with the advent of 5G and beyond, there is an increasing shift toward distributed, more adaptive, and autonomous solutions, such as Reinforcement Learning (RL).

In the context of RRM in wireless communication systems, \textit{single-agent} RL is typically used in scenarios where \textit{centralized} approaches are feasible and decision-making can be isolated within one entity, such as in a single base station. For example, in \cite{zhang2019deep}, deep Q-learning is employed to train a single agent to handle all decisions regarding user association, resource allocation, and power allocation. 
In \cite{he2019joint}, policy gradient method is applied to the Attention-based network architecture to create an RL agent, which is used for channel assignment at a central base station of a multi-carrier NOMA system. In addressing spectrum resource scarcity in Industrial Internet of Things networks, Q-learning is employed in \cite{shi2020deep} to facilitate dynamic spectrum resource management and enable spectrum sharing between different types of user equipment.

On the other hand, \textit{multi-agent} RL is more applicable in \textit{decentralized} systems that involve distributed entities where deployment of a single central controller is impossible due to unknown interfering channels, signaling overhead or delays. In such systems, multiple agents must collaborate and learn to use resources in a coordinated manner. Studies like \cite{tang2020deep,  zhao2019deep,yang2022distributed, zhang2020deep, mei2024multi} 
demonstrated MARL for radio resource management in HetNets, where multiple agents are built to cooperatively perform channel assignment and power allocation. Similarly, MARL has been shown to be effective for managing spectrum sharing and transmit power in highly dynamic communication systems such as vehicular-to-vehicular (V2V) \cite{li2022federated, liang2019spectrum, ji2023multi}, cellular vehicular-to-everything (C-V2X)  \cite{hegde2023radio, vu2022multi, guo2023radio, azizi2024efficient}. Closer to our work are studies that employ MARL for resource management in traditional cellular systems  \cite{naderializadeh2021resource, nasir2019multi,Yang24TWC:DRLMixedDataset}. In these studies, an independent learning mechanism is adopted to build multi-agent systems, where each agent is trained to handle either user scheduling or power allocation tasks. To avoid action space explosion, these works restrict to time-division duplex (TDD) usage, or a limited number of transmit power steps. A recent study in \cite{ji2024meta} proposes a framework which combines the ideas of meta-learning and federated learning to address the channel assignment and transmit power optimization problem. Specifically, PPO algorithm is used to train a pre-trained model at the base station. This pre-trained model is then broadcasted to local users to perform resource allocation and for further training whenever a new user joints the network. Users then send back their fine-tuned models to the base station to update the pre-trained model by averaging. The action space explosion is avoided by imposing the restriction that each user can be assigned at most one subchannel. Although the authors claim to address the issue of user dynamics, migrating training to local users is unrealistic due to user's limited computation.


To sum up, the effectiveness of RL and MARL and their potentials in RRM is undeniable. However, these approaches face a critical challenge: the exponential explosion of the action space, which prevents solutions from scaling to practical configurations and generalizing to the dynamics of real-world systems. Existing studies either exclude spectrum management to avoid this challenge \cite{naderializadeh2021resource} or limit themselves to simulations on toy configurations with a small number of subchannels \cite{tang2020deep, zhao2019deep, yang2022distributed}. This study is dedicated to filling this gap.

\section{System model and problem formulations} 
\label{sec:system}
Consider a downlink system with a set $\mathcal{N}$ of $N$ base stations, possibly from TN or NTN segments, serving a set of ground users via a shared frequency bandwidth. We consider practical scenarios with the user set varying over time. Although this consideration makes no difference in conventional time slot-based optimization solutions, the evolution of the user set has a significant impact in learning-based solutions, as it directly affects the dynamics of the learning environment. Formally, let $\mathcal{K}^t$ be the set of active $K^t$ users in the system at time step $t$ and $\mathcal{K}^t_n \subseteq \mathcal{K}^t$ be the subset of users connected with BS $n$ that is non-overlapped with user subsets of other BSs. 

The BSs serve their single-antenna users through OFDM transmission to mitigate intra-cell interference, in which the total bandwidth $B$ is divided into the set $\mathcal{F}$ of $F$ subchannels, each with a bandwidth of $W = B/F$. Different numerologies will result in different number of the total subchannels and subcarrier spacings \cite{3GPP5GNR}. 
Inter-cell interference will be managed via proper subchannel assignment and power allocation. 
Particularly, let $x^t_{k,n, f}\in \{0, 1\}$ be a binary variable indicating whether the subchannel $f$ of the BS $n$ is allocated to user $k \in \mathcal{K}^t_n$ at the time step $t$. Let $\mathbf{x}^t$ be the accumulation of all these binary variables. The orthogonal constraint at this time step is then represented by \begin{equation}
    {\sum}_{k\in\mathcal{K}^t_n}x^t_{k,n,f} \leq 1, \forall f\in\mathcal{F}, n\in\mathcal{N}. \label{ctr:orthogonal}
\end{equation}
Let $\boldsymbol{h}^t_{k,n,f}\in \mathcal{C}^{n_T\times 1}$ be the channel coefficients between the $n_T$ antennas of BS $n$ and user $k$ on subchannel $f$ at the time step $t$. To maximize the receive signal to noise ratio, maximum ratio transmission (MRT) is used on every subchannel. As a result, the SINR at user $k$ of BS $n$ with channel assignment $\mathbf{x}^t \triangleq \{x^t_{k,n,f}\}_{\forall k,n,f}$ is given as:
\begin{equation}
    \gamma^t_{k, n, f}(\mathbf{x}^t)\! = \!\frac{|{\boldsymbol{h}}^t_{k, n,f}|^2 p^t_{k,n,f} x^t_{k,n,f}}{\underset{l\in\mathcal{N},l\neq n}{\sum}\ \underset{k'\in\mathcal{K}_l}{\sum}|\boldsymbol{h}^t_{k,l,f}|^2 p^t_{k',l,f} x^t_{k',l,f}\! +\! WN_0} \label{eq:sinr}
\end{equation}
where $p^t_{k,n, f}$ is power scaling factor for user $k$ from BS $n$ at time step $t$. In \eqref{eq:sinr}, the first term in the denominator represents  the aggregated interference caused by other BSs' transmission on the same subchannel. Note that intra-cell interference is mitigated by the OFDM transmission constraint in \eqref{ctr:orthogonal}. 
The achievable rate at the time step $t$ for user $k$ on the subchannel $f$ under the assignment $\mathbf{x}^t$ is then given by
\begin{equation}
    R^t_{k, n, f}(\mathbf{x}^t) = W\log_2\left(1 + \gamma^t_{k, n, f}(\mathbf{x}^t)\right). \label{eq:rate}
\end{equation}

\subsection{Problem Formulation and Optimization-based Solution}
We aim to maximize the whole system sum-rate, subject to users' QoS requirements $\eta^t_{k,n}$ at the time step $t$, which can then be formulated as
\begin{align}
    \max_{\textbf{x}^t, \boldsymbol{p}^t}  \quad& \sum_{n\in\mathcal{N}}\sum_{k\in\mathcal{K}^t_n}\sum_{f\in \mathcal{F}}R^t_{k, n, f}(\mathbf{x}^t)\label{prob:max_sum_rate}\tag{\sf P1} \\
    \text{s.t.} \quad
    &{\sum}_{f\in \mathcal{F}}R^t_{k, n, f}(\mathbf{x}^t) \geq \eta^t_{k,n}, \forall k\in \mathcal{K}^t_n, n\in \mathcal{N}\label{ctr:qos}\\
    &{\sum}_{k\in\mathcal{K}^t_n}x^t_{k,n,f} \leq 1, \forall f\in\mathcal{F}, n\in\mathcal{N}\\
    &x^t_{k,n,f} \in \{0, 1\}, \forall k\in \mathcal{K}^t_n, f\in\mathcal{F}, n\in\mathcal{N}\\
    & {\sum}_{f\in \mathcal{F}}{\sum}_{k \in \mathcal{K}^t_n} p^t_{k,n,f} \leq p_{max}, \forall n \in\mathcal{N}
\end{align}
where $p_{max}$ is the total transmit power at every BS and $\boldsymbol{p}^t$ is a short-hand notation for all $\{p^t_{k,n,f}\}_{\forall k,n,f}$. Problem \eqref{prob:max_sum_rate} represents a fundamental challenge in practical wireless communication systems. Indeed, solving \eqref{prob:max_sum_rate} by centralized optimization \cite{FatemehTWC25} is impractical for real-world systems due to i) the lack of global CSI and ii) excessive signaling overhead and latency. In practice, the 3GPP standardization only allows a BS to estimate CSI to its connected users, making it is impossible to solve problem \eqref{prob:max_sum_rate} optimally due to the lack of interference channel coefficients.

An alternative solution is to decompose~\eqref{prob:max_sum_rate} into $N$ subproblems and independently solve each subproblem at the corresponding BS. However, since each BS cannot estimate the interference channel coefficients from other BSs, i.e., $\boldsymbol{h^t}_{k,n,f}, \forall k \notin \mathcal{K}^t_n$, the exact SINR in \eqref{eq:sinr} cannot be used. Instead, each BS can approximate the aggregated interference from other BSs using the interference channel statistic, and assume uniform power allocation at all BSs. Thus, the approximated SINR can be written as  
\begin{equation}
    \Tilde{\gamma}^t_{k,n,f}(\mathbf{x}^t_n) = \frac{|{h}^t_{k, n,f}|^2p{x}^t_{k,n,f}}{{\sum}_{l\in\mathcal{N},l\neq n} L^t_{k, l}p + WN_0}
\end{equation}
where $p=\frac{p_{max}}{F}$ is the uniform transmit power per subchannel and $L^t_{k, l}$ is the path loss between BS $l$ and user $k$ of BS $n$ at time step $t$. Here, we ignored the effects of fast-fading on the interference channels and assumed all subchannels at other BSs are activated. Let $\Tilde{R}^t_{k, n, f} = W\log_2\left(1 + \Tilde{\gamma}^t_{k,n,f}(\mathbf{x}^t_n)\right)$ be the approximate data rate achieved when the subchannel $f$ is assigned to user $k$ (i.e., $x^t_{k,n,f}$ is set to 1). Then the subproblem to be solved at BS $n$ at time step $t$ reads as follows:
\begin{align}
    \max_{\textbf{x}^t_n}  \quad& {\sum}_{k\in\mathcal{K}^t_n}{\sum}_{f\in \mathcal{F}}\Tilde{R}^t_{k, n, f}\label{prob:max_sum_rate_approx}\tag{\sf P2} \\
    \text{s.t.} \quad
    &{\sum}_{f\in \mathcal{F}}\Tilde{R}^t_{k, n, f} \geq \eta^t_{k,n}, \forall k\in \mathcal{K}^t_n\\
    &{\sum}_{k\in\mathcal{K}^t_n}x^t_{k,n,f} \leq 1, \forall f\in\mathcal{F}\\
    &x^t_{k,n,f} \in \{0, 1\}, \forall k\in \mathcal{K}^t_n, f\in\mathcal{F}.
\end{align}
Problem~\eqref{prob:max_sum_rate_approx}
is an integer linear program, where approximate solution can be obtained easily by a linear relaxation. However, even if there is an effective procedure to solve \eqref{prob:max_sum_rate_approx} in an exact manner, this decentralized optimization approach has significant drawbacks. First, the QoS constraints in \eqref{ctr:qos} may be violated due to errors in interference approximation. Second, since resource allocations are handled separately at different BSs without inter-BS coordination, there is no cooperative mechanism to efficiently mitigate interference. As a result, selfish resource allocation at one BS may negatively affect the strategies of other BSs. 

\subsection{Standard Constrained Dec-POMDP Formulation}
To make MARL applicable in this context, the problem needs to be reformulated as a decentralized partially observable Markov decision process (Dec-POMDP). The most common and straightforward approach is to model each BS as an agent \cite{tang2020deep,naderializadeh2021resource,mei2024multi}, responsible for subchannel assignment of that BS, assuming equal power allocation. In particular, problem \eqref{prob:max_sum_rate} can be represented by a tuple $\langle\mathcal{N}, \mathcal{S}, \mathcal{A}, P, r, \mathcal{Z}, \gamma\rangle$, where $\mathcal{N}$ is the set of $N$ agents, $\mathcal{S}$ is the state space, $\mathcal{A} = \prod_{n\in\mathcal{N}}\mathcal{A}_{n}$ is the joint action space with $\mathcal{A}_{n}$ denoting the action space of agent $n$, $P:\mathcal{S}\times\mathcal{A}\times\mathcal{S}\rightarrow\mathbb{R}$ is the environment transition kernel where $P(s'|a, s)$ is the probability of moving to the next state $s'$ after taking action $a$ in state $s$, $r:\mathcal{S}\times\mathcal{A}\rightarrow\mathbb{R}$ is the reward function shared by all agents, $\mathcal{Z} = \prod_{n\in\mathcal{N}}\mathcal{Z}_{n}$ is the joint observation space, and $\gamma$ is the discount factor.

At time step $t$, let $a^t_{n}\in \mathcal{A}_{n}$ be the action of agent $n$ and $a^t\in \mathcal{A}$ be the joint action of all agents. Similarly, let $z^t_{n}\in\mathcal{Z}_{n}$ be the observation of agent $n$ and $s^t\in\mathcal{S}$ be the global state.
Let $H^t_{n} \in (\mathcal{Z}_n\times\mathcal{A}_n)^{t-1}\times \mathcal{Z}_n$ be the action-observation history of agent $n$ at the time step $t$ given by $H^t_{n} = (z^0_{n}, a^0_{n}, z^1_{n}, \ldots, a^{t-1}_{n}, z^t_{n})$.
Let $r(s^t, a^t)$ be the reward for taking the joint action $a^t$ in state $s^t$ and $\pi_{n}:(\mathcal{Z}_n\times\mathcal{A}_n)^{t-1}\times \mathcal{Z}_n\rightarrow \Delta(\mathcal{A}_n)$ be the policy of agent $n$ at BS $n$, where $\Delta(\mathcal{A}_n)$ is a probability simplex over the action space $\mathcal{A}_n$, which represents the probability of assigning the sub-channels to active users of agent $\mathtt{A}_n$.
We are to find a joint policy $\pi$, where $\pi(a^t|H^t_1, \ldots, H^t_N) = \prod_{n\in\mathcal{N}}\pi_n(a^t_{n}|H^t_n)$, to maximize the expectation of total discounted reward while satisfying QoS constraints:
\begin{align}
    &\underset{\pi}{\text{max}}\quad  \mathbb{E}_{ \pi}\left[{\sum}_{t=0}^{\infty}\gamma^tr(s^t, a^t)\right]\label{prob:cmdp}\tag{\sf RL1}\\
    &\text{s.t.}\ 
    {\sum}_{f\in \mathcal{F}}R^t_{k, n, f}(a^t) \geq \eta_{k,n}, \forall k\in \mathcal{K}^t_n, n\in \mathcal{N},t\in\mathbb{N}\label{ctr:cmdp},
\end{align}
where $R^t_{k, n, f}(a^t)$ denotes the achievable rate for user $k$ on the subchannel $f$ of BS $n$, induced by the joint action $a^t$ of all agents at time step $t$.

The formulation in \eqref{prob:cmdp} is standard and it can be solved by applying conventional MARL frameworks for small problem sizes. This is because learning a joint policy requires a combinatorial complexity of the joint action space. Such combinatorial complexity is prohibited in practical scenarios with a large number of subchannels and users. For example, with 3 BSs operating a 20MHz bandwidth and 120kHz numerology, each serving 10 users, the joint action space cardinality is about 4 billion, which is practically impossible to learn. Moreover, although constraint~\eqref{ctr:cmdp} can be handled via Lagrangian relaxation, this approach significantly increases training complexity and often impedes convergence, even under simplified conditions where all users share identical QoS requirements at all time steps. In our practical setting, users with heterogeneous QoS requirements can join or leave the network unpredictably, introducing substantial environmental variability. In addition, MARL inherently suffers from nonstationarity, as the concurrent policy updates of agents dynamically alter each other's learning environment. The interplay between external dynamics and internal nonstationarity results in an unstable training process, posing a major challenge to learning effective and reliable policies. In this work, we propose a new method for handling the QoS constraints and a scalable MARL algorithm to avoid the exponential action space of conventional MARL solutions. 

\section{Proposed Framework}
\label{sec:solution}
As analyzed in previous section, conventional MARL solutions struggle with \emph{scalability} and \emph{constraint satisfaction} of the RRM with medium- and large-scale problem sizes. We tackle these challenges by proposing a hybrid model-learning framework for distributed RRM in 6G systems. We start by reviewing the motivation of our design and the high-level idea of the framework. After that, each phase of the proposed framework is presented in detail.

\subsection{Methodological Overview}
Conventional MARL approaches struggle with large problem sizes for two main reasons. First, directly learning constraint satisfaction is notoriously inefficient, especially in a large-scale system with a stochastic set of users and QoS demands \cite{Wachi20:ICML:safeRL}. Second, the joint action space grows exponentially with the number of subchannels, making direct policy learning computationally intractable and sample-inefficient. These two challenges motivate our hybrid learning framework, in which the decision-making process comprises two phases as follows:

\begin{itemize}
    \item \textbf{Phase I} - Greedy Constraint Satisfaction: The first phase of our framework is designed to allocate a minimum number of subchannels to deterministically satisfy the users' QoS requirements, preserving resources for the optimization in the second phase. 
    By offloading the responsibility of constraint satisfaction to this greedy procedure, we address the first challenge of constraint handling, enhance reliability of the solution, and significantly reduce the burden on the learning process for the second phase. This phase serves as the expert in RLHF \cite{kaufmann2024RLHF} dealing with the most difficult task of constraint satisfaction. 
    \item \textbf{Phase II} – Scalable MARL-Based Sum-Rate Maximization: we train a multi-agent system to optimally allocate the remaining available subchannels to maximize the overall system throughput. To address the second challenge of action space explosion towards a scalable learning solution, we exploit the interference property of the OFDM transmission: (inter-cell) interference on a particular subchannel $f$ is only caused by transmissions from other BSs on the same subchannel $f$. This key property allows us to decompose the global MARL problem into independent subchannel-wise subproblems. Each of these subproblems is solved using a separate instance of QMIX \cite{rashid2020monotonic,Cuong24UAV}, enabling scalable and stable learning. More importantly, since the QoS constraints have already been satisfied in the first phase, the second phase only involves unconstrained MARL problems focused purely on maximizing system throughput. This not only reduces the burden on the learning but also improves the convergence speed, training stability, and overall solution quality.
\end{itemize}
\begin{figure}
    \centering
    \includegraphics[width=0.99\linewidth]{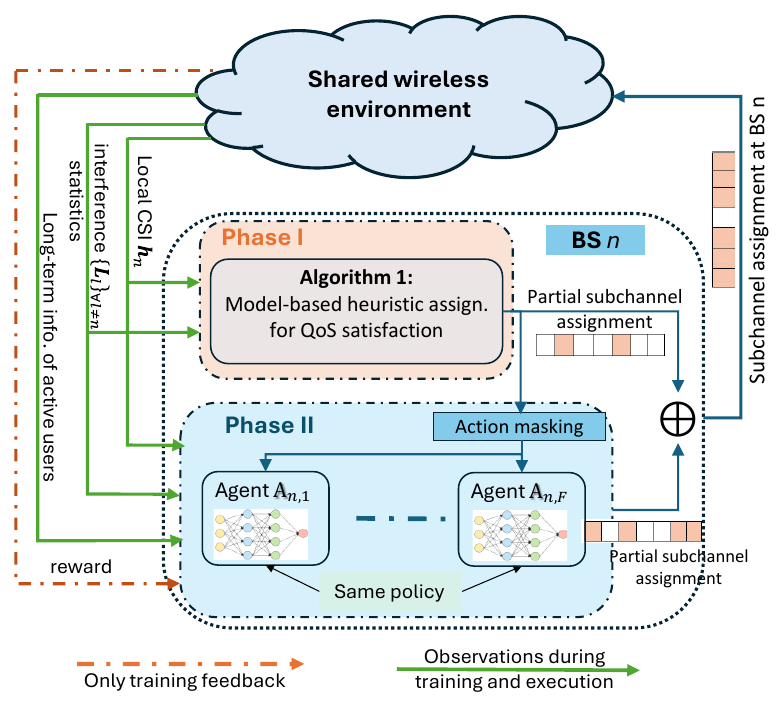}
    \caption{Block-diagram of the proposed hybrid learning framework. Each BS only observes local CSIs and long-term network information.}
    \label{fig:greq_architecture}
    \vspace{-0.5cm}
\end{figure}
Fig.~\ref{fig:greq_architecture} presents the block-digram of our proposed framework. 

\subsection{Phase I: Greedy Assignment for QoS Satisfaction}
\begin{algorithm}[!b]
\caption{Greedy Assignment for QoS Satisfaction}
\label{alg:greedy-qos}
\begin{algorithmic}[1]
\renewcommand{\algorithmicrequire}{\textbf{Input:}}
\renewcommand{\algorithmicensure}{\textbf{Output:}}
\REQUIRE Local channel gains and rate requirements of users associated with BS $n$, pathloss.
\ENSURE Greedy channel assignment $S_n$ that satisfies the QoS requirements at BS $n$.
\STATE ${S}^t_{n,f} \gets \infty$ for all $f\in \mathcal{F}$;
\STATE Initialize the approximate accumulated data rate $\rho^t_{k,n} \gets 0$ for each user $k\in \mathcal{K}^t_n$;
\STATE $\mathcal{D} \gets$ set of available subchannels;
\REPEAT
    \FOR{each user $k\in \mathcal{K}^t_n$}

        \IF{$\rho^t_{k,n} \geq (1 + \epsilon)\eta^t_{k,n}$}
            \STATE Skip this user;
        \ELSE
        \STATE $\Tilde{\gamma}^t_{k,n,f} \gets \frac{|{h}^t_{k, n,f}|^2p}{\underset{l\in\mathcal{N},l\neq n}{\sum}L^t_{k, l}p + WN_0}$;
        \STATE $f \gets \underset{f\in \mathcal{D}}{\arg\max}~\Tilde{\gamma}^t_{k,n,f}$;
        \STATE Assign subchannel $f$ to user $k$: $S^t_{n, f} \gets k$;
        \STATE $\rho^t_{k,n} \gets \rho^t_{k,n} + W\log_2(1 + \Tilde{\gamma}^t_{k,n,f})$;
        \STATE $\mathcal{D}\gets\mathcal{D}\backslash \{f\}$;
        \ENDIF
    \ENDFOR
\UNTIL{$\rho^t_{k,n} \geq (1 + \epsilon)\eta^t_{k,n}$ for all $k\in\mathcal{K}^t_n$ OR $\mathcal{D} = \varnothing$;}
\RETURN $S_n$.
\end{algorithmic}
\end{algorithm}

The greedy channel assignment at the BS $n$ is outlined in Algorithm \ref{alg:greedy-qos}. In particular, the output $S_n$ is a vector of size $F$  representing the partially filled channel assignments, where $S_{n,f} = k$ indicates that subchannel $f$ is assigned to user $k\in \mathcal{K}^t_n$ and $S_{n,f} = \infty$ indicates that this subchannel remains unassigned.
Our objective is to approximately satisfy user QoS requirements using as few subchannels as possible. To achieve this, the algorithm needs to track the cumulative data rate of each user during the greedy allocation. However, since each BS cannot access the interference channels or the decisions of other BSs in real time, the cumulative data rate can only be approximated by $\rho^t_{k,n}$, based on the pathloss of the interference channels and the assumption that all subchannels at other BSs are active.
In particular, if the subchannel $f$ of BS $n$ is assigned to the user $k$, the rate increment is approximated by $W \log_2\left(1 + \frac{|\boldsymbol{h}_{k,n,f}|^2 p}{\sum_{l \ne n} L_{k,l} p + WN_0}\right)$, which is added to the user $k$’s cumulative rate $\rho^t_{k,n}$. 


Using the above indicator, at each time step, every BS executes the greedy assignment independently using only its local CSI and the interference channels statistics, i.e., pathloss. The procedure in Algorithm \ref{alg:greedy-qos} proceeds iteratively where in each round, it checks the current accumulative rate of each associated user (line 6). If a user's QoS target $\eta^t_{k,n}$ is met, the user is removed from further consideration (line 7). Otherwise, the user is assigned the subchannel with the highest approximated SINR (lines 9-11). The user’s accumulative rate is then updated (line 12), and the assigned subchannel is removed from the set of available subchannels (line 13). This process is repeated until either all users meet their QoS or no subchannels remain (line 16). Note that since $\rho^t_{n,k}$ is only an approximation of the actual cumulative data rate, there's a chance the true rate may fall short of meeting the QoS requirement. This can happen due to constructive fading on the interference channels, which was not accounted for in the approximation. To reduce the risk of this happening, a safety margin ratio $\epsilon$ is added to the QoS requirements, as shown in the conditions in lines 6 and 16 of Algorithm \ref{alg:greedy-qos}. In general, the value of $\epsilon$ can be estimated by integrating over the distribution of the interference channel coefficients. We numerically found that an $\epsilon \simeq 3\%$ achieves a good trade-off between the QoS satisfaction and the system sum rate. Mathematical analysis on $\epsilon$ selection is not considered in the current work.


\subsection{Phase II: Scalable MARL-Based Sum-rate Maximization}
We first show how to decompose the problem into smaller, tractable subproblems. Then, given the output of the greedy assignment procedure in Algorithm~\ref{alg:greedy-qos}, we will describe how to integrate this procedure into the MARL framework to train the agents used in phase II of the proposed framework.

\subsubsection{Problem decomposition}
Our goal is to decompose problem~\eqref{prob:cmdp} into multiple subproblems that are easier to solve, while ensuring that the decomposition does not incur performance loss. To this end, rather than training a single agent for each BS to manage all $F$ subchannels jointly, which leads to an intractably large action space, we adopt a fine-grained agent design in which each BS is equipped with $F$ agents, one per subchannel. Specifically, denoting $\mathtt{A}_{n,f}$ as the agent responsible for assigning the $f$th subchannel at BS $n$ and $\mathcal{G}_f = \{\mathtt{A}_{1,f}, \mathtt{A}_{2,f}, \ldots, \mathtt{A}_{N,f}\}$ as the group of agents managing subchannel $f$ across all $N$ BSs. This design exploits interference-free across different subchannel indexes of OFDM systems. Consequently, the $F$ agents at each BS can be fully decoupled, i.e., agent $\mathtt{A}_{n,f}$ is independent from other agents $\mathtt{A}_{n,f'}, \forall f' \neq f$. In fact, agent $\mathtt{A}_{n,f}$ needs to coordinate only with other $N-1$ agents of $\mathcal{G}_f$ to mitigate inter-cell interference, which avoids unnecessary interactions between agents within one BS, thereby reducing the nonstationarity faced by each agent during training. 
The $N$ agents in each group $\mathcal{G}_f$ share a common reward function, encouraging them to align their actions to mitigate interference while maximizing the sum rate. At a high level, the large-scale MARL problem \eqref{prob:cmdp} is decomposed into $F$ subproblems, where the $f$th subproblem involves training the group $\mathcal{G}_f$ of $N$ agents. Importantly, each agent has a much smaller action and state space.

Following the above idea, the $f$th subproblem can be presented by the tuple $\langle \mathcal{G}_f, \mathcal{S}_f, \mathcal{A}_f, P_f, r_f, \mathcal{Z}_f, \gamma \rangle$, where 
$\mathcal{S}_f$ is the state space of subchannel $f$;
$\mathcal{A}_f = \prod_{n\in\mathcal{N}} \mathcal{A}_{n,f}$ is the joint action space, with $\mathcal{A}_{n,f}$ denoting the action space of agent $\mathtt{A}_{n,f}$;
$P_f:\mathcal{S}_f\times\mathcal{A}_f\times\mathcal{S}_f\rightarrow\mathbb{R}$ is the environment transition kernel where $P_f(s'|a, s)$ is the probability of moving to the next state $s'$ after taking action $a$ in state $s$;
$r_f: \mathcal{S}_f \times \mathcal{A}_{f} \rightarrow \mathbb{R}$ is the reward function shared by $N$ agents of group $\mathcal{G}_f$; $\mathcal{Z}_f = \prod_{n \in \mathcal{N}} \mathcal{Z}_{n,f}$ is the joint observation space, where $\mathcal{Z}_{n,f}$ is the observation space of agent $\mathtt{A}_{n,f}$; and $\gamma$ is the discount factor of the learning environment. 
At each time step $t$, let $a^t_{n,f} \in \mathcal{A}_{n,f}$ be the action taken by agent $\mathtt{A}_{n,f}$, and $a^t_f = (a^t_{1,f}, \ldots, a^t_{N,f})$ denote the joint action of group $\mathcal{G}_f$. Let $z^t_{n,f} \in \mathcal{Z}_{n,f}$ be the local observation of agent $\mathtt{A}_{n,f}$ and $s^t_f \in \mathcal{S}_f$ be the global state. Each agent maintains an action-observation history $H^t_{n,f} = (z^0_{n,f}, a^0_{n,f}, \ldots, a^{t-1}_{n,f}, z^t_{n,f}) \in (\mathcal{Z}_{n,f} \times \mathcal{A}_{n,f})^{t-1} \times \mathcal{Z}_{n,f}$. Let $\pi_{n,f}: (\mathcal{Z}_{n,f} \times \mathcal{A}_{n,f})^{t-1} \times \mathcal{Z}_{n,f} \rightarrow \Delta(\mathcal{A}_{n,f})$ denote the policy of agent $\mathtt{A}_{n,f}$, where $\Delta(\mathcal{A}_{n,f})$ is the probability simplex over its action space. The goal of the $f$th subproblem is to learn a joint policy $\pi_f = \prod_{n \in \mathcal{N}} \pi_{n,f}$ that maximizes the expected discounted return under group reward $r_f$:
\begin{align}
    \underset{\pi_f}{\text{max}} \quad &\mathbb{E}_{\pi_f} \left[ {\sum}_{t=0}^{\infty} \gamma^t r_f(s^t_f, a^t_f) \right]. \label{prob:cmdp_decomposed}\tag{\sf RL2}
\end{align}


\subsubsection{Details of the learning environment}
\label{subsec:env}
To handle the dynamics of the user set, we allow any number of active users per BS that does not exceed  $K_{\max}$ - the maximum number of users can be simultaneously served. Following this, each BS has $K_{\max}$ serving slots. In practice, the value of $K_{\max}$ is determined by the total bandwidth $B$ and the services. When a new user joins the system, it is assigned to one of the idle slots. If all $K_{\max}$ slots are occupied, the user is denied service. Conversely, when a user leaves the network, its corresponding slot is marked as idle and becomes available for future assignments. This assumption reflects practical constraints, as real-world wireless systems operate under limited resources and can only support a bounded number of users concurrently. Under this model, we can view the resource allocation problem as assigning resources to the fixed set of serving slots rather than assigning resources to a dynamic set of users. Based on this abstraction, the components of the learning environment can now be specified as follows:
\begin{itemize}
    \item \textit{Action space}: the action space of agent $\mathtt{A}_{n,f}$ is defined as $\mathcal{A}_{n,f} = \{0, 1, \ldots, K_{\max}\}$, where $a^t_{n,f} = \alpha \in \mathcal{A}_{n,f}$ indicates that the $f$th subchannel of BS $n$ is assigned to the $\alpha$th serving slot of this BS. Notably, the additional action $0$ in this definition corresponds to the \emph{unactivated} action, meaning that subchannel $f$ is not assigned to any slot during time step $t$. Under the shared reward function, including this unactivated action has significant impact, as it enables agents to collaboratively decide to deactivate certain subchannels for \emph{interference mitigation}.
    \item \textit{Invalid action masking}: to ensure that the second phase does not override the assignments made in the first phase and that resources are not allocated to idle serving slots, we apply \emph{invalid action masking} as follows. If the $f$th subchannel of BS $n$ has already been assigned in the first phase of this allocation step, i.e., $S^t_{n,f} \neq \infty$, then all actions except $S^t_{n,f}$ are masked out (equivalent to not calling agent $\mathtt{A}_{n,f}$ in this step), thereby forcing the agent to preserve the previous decision. Otherwise, all actions corresponding to idle serving slots are masked out, restricting the agent to either deactivate the subchannel (i.e., select action $0$) or assign it to one of the currently occupied slots. With this design, we effectively transform the original value-based constraint in \eqref{ctr:cmdp} into a constraint on the agent's feasible action set. This shift greatly simplifies the training process, as it avoids the need for complex constraint-handling mechanisms. Instead, the constraint can be enforced directly through invalid action masking, a lightweight and widely used technique in the RL literature \cite{Huang22ActionMasking}. This not only improves training stability but also makes the implementation more reliable and scalable.
    \item \textit{Reward function}: The shared reward function for agent group $\mathcal{G}_f$ is defined as the sum rate achieved across all BSs on subchannel $f$ as follows:
    \begin{equation}
        r_{f}(s^t_f, a^t_f) = {\sum}_{n\in\mathcal{N}} {\sum}_{k\in\mathcal{K}^t_n}R^t_{k,n,f}(a^t_f),
    \end{equation}
    where $R_{k,n,f}(a^t_f)$ denotes the achievable data rate on subchannel $f$ of user $k$ served by BS $n$, resulting from the joint action $a^t_f$. 
    \item \textit{Observation}: Observation $z^t_{n,f}$ of agent $\mathtt{A}_{n,f}$ at time $t$ consists of local channel gains of BS $n$'s users on subchannel $f$ normalized by the noise power and  total interference power experienced by the users of BS $n$. In practice, the total interference can be extracted from Reference Signal Strength Indicator (RSSI) report \cite{3GPP5GNR}. Additionally, we assume available long-term system information which is also inputted to the agent, including network-wise user locations and the number of active users in all BSs. This practical and local observation model enables a \emph{fully distributed deployment} of the trained agents at every BS.

    \item \textit{State space}: The state $s^t_f$ of the $f$th subproblem at time $t$ comprises the full channel gains of the entire system on subchannel $f$, along with a vector encoding the occupancy status of serving slots at every BS.
\end{itemize}
With the above design, instead of training a single agent per BS with an exponentially large action space of size $K_n^F$, each BS is controlled by $F$ agents operating independently, each having an action space of size $K_{\max} + 1$. Moreover, by leveraging the interference structure of the OFDM system, the optimal actions of agents are jointly optimal for the original problem, thereby avoiding any performance loss due to decomposition. Technically, all agents can share the same policy by incorporating the subchannel index $f$ and the BS index $n$ into their observations, thus requiring training of only one policy for the whole system.

The overall training procedure is summarized in Algorithm~\ref{alg:learn}, where the greedy assignment from Algorithm \ref{alg:greedy-qos} is invoked at every decision step to ensure QoS requirements are met.

\begin{algorithm}[t]
\caption{Integrated Training Procedure}
\label{alg:learn}
\begin{algorithmic}[1]
\renewcommand{\algorithmicrequire}{\textbf{Input:}}
\renewcommand{\algorithmicensure}{\textbf{Output:}}
\REQUIRE $\mathsf{num\_episodes}$.
\ENSURE Trained policies.
\STATE $\mathsf{episodes\_done} \gets 0$;
\WHILE{$\mathsf{episodes\_done} < \mathsf{num\_episodes}$}
    \STATE Reset environment;
    \FOR{each step $t$ of the episode}
        \FOR{each base station $n$}
            \STATE $S^t_n\gets$ Greedy assignment; \tcp{Algorithm \ref{alg:greedy-qos}}
            \FOR{each subchannel $f$}
                \IF{$S^t_{n,f} = \infty$}
                    \STATE $S^t_{n,f}\gets$ assign subchannel $f$ by calling agent $\mathtt{A}_{n,f}$;
                \ENDIF
            \ENDFOR
        \ENDFOR
        \STATE Commit the assignment $\mathbf{S}^t = \{S^t_1,\ldots,S^t_N\}$ and obtain reward;
        \STATE Update policies;
    \ENDFOR
    \STATE $\mathsf{episodes\_done} \gets \mathsf{episodes\_done} + 1$;
\ENDWHILE
\end{algorithmic}
\end{algorithm}

\subsubsection{QMIX algorithm for training agents}
\begin{figure}
    \centering
    \includegraphics[width=0.99\linewidth]{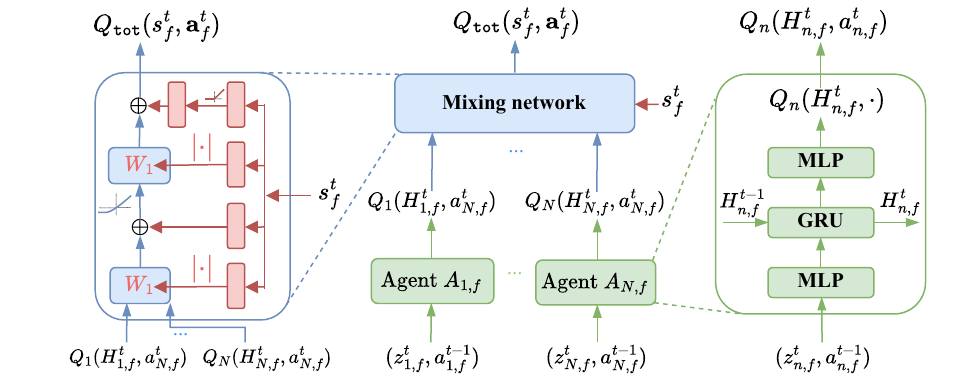}
    \caption{The detailed architectures of the mixing network (blue) and agent networks (green).}
    \label{fig:qmix}
\end{figure}

For each agent group $\mathcal{G}_f$, we employ the QMIX algorithm~\cite{rashid2020monotonic,Cuong24UAV} to enable centralized training and decentralized execution. QMIX models a global action-value function $Q_{\texttt{tot}}(s^t_f, a^t_f)$, which evaluates the joint action $a^t_f$ taken in the state $s^t_f$. This joint value function is factorized into individual utility functions $ Q_n(H^t_{n,f}, a^t_{n,f}) $, where each function estimates the value of a local action $ a^t_{n,f} $ given the agent’s local action-observation history $ H^t_{n,f} $. A key property of QMIX is that it enforces a monotonic relationship between the joint value function and the individual utilities:
\[
\frac{\partial Q_{\texttt{tot}}}{\partial Q_n} \geq 0, \quad \forall n \in \mathcal{N},
\]
ensuring that increasing an agent’s local utility never decreases the global value. This monotonicity allows the joint greedy action to be recovered by independently selecting the local greedy actions as follows:
\begin{equation}
    \underset{a^t_f}{\arg\max}~Q_{\texttt{tot}}\!(\!s^t_f, a^t_f\!)\! =\! \left(\!\begin{matrix}
                        \underset{a^t_{1,f}}{\arg\max}~Q_1\!(\!H^t_{1,f}, a^t_{1,f})\\
                        \vdots\\\underset{a^t_{N,f}}{\arg\max}~Q_N\!(\! H^t_{N,f}, a^t_{N,f})
                    \end{matrix}\!\right).
\end{equation}
This decomposition enables decentralized decision-making at execution time, where each agent selects actions using only its local utility function $Q_n$, while training is guided by a global critic $Q_{\texttt{tot}}$ that captures joint dependencies. As a result, QMIX facilitates scalable multi-agent learning without sacrificing coordination performance.

Fig.~\ref{fig:qmix} illustrates the neural architecture adopted in our framework. It consists of two key components: (i)~\emph{agent networks}, which compute individual action-value functions $ Q_n(H^t_{n,f}, a^t_{n,f}) $ based on each agent’s local action-observation history; and (ii)~a \emph{mixing network}, which aggregates the individual $ Q_n $ values into a joint action-value function $ Q_{\texttt{tot}}(s^t_f, a^t_f) $, conditioned on the global state $ s^t_f $. Each agent network is implemented as a Deep Recurrent Q-Network (DRQN)~\cite{hausknecht2015deep} with parameter sharing applied across all agents to improve sample efficiency and promote generalization. The mixing network is realized as a two-layer feedforward neural network with Exponential Linear Unit (ELU) activations. To preserve the monotonicity condition required by QMIX, i.e., $ \partial Q_{\texttt{tot}}/\partial Q_n \geq 0 $, the weights of the mixing network are constrained to be non-negative. This is achieved by using a \emph{hypernetwork} that generates the weights and biases of the mixing network as functions of the global state. The hypernetwork consists of fully connected layers with ReLU activations, followed by element-wise absolute operations to ensure non-negativity.

The entire framework is trained end-to-end using experience replay. The objective is to minimize the temporal-difference (TD) loss between the predicted joint action-value and a one-step bootstrapped target 
%
$\mathcal{L}(\theta^t) = \mathbb{E}\big[\big(y_{\texttt{tot}} - Q_{\texttt{tot}}(s^t_f, a^t_f; \theta^t)\big)^2\big]$,
where $ \theta^t $ denotes the parameters of the online (primary) network, and the target value $ y_{\texttt{tot}} $ is computed as
\begin{equation}
    y_{\texttt{tot}} = r_f(s^t_f, a^t_f) + \gamma {\max}_{a^{t+1}_f} Q_{\texttt{tot}}(s^{t+1}_f, a^{t+1}_f; \Bar{\theta}^t),
\end{equation}
with $ \Bar{\theta}^t $ representing the parameters of the slowly updated target network. The replay buffer stores transitions of the form $(s^t_f, z^t_f, a^t_f, r_f, s^{t+1}_f, z^{t+1}_f)$, from which mini-batches are sampled during training.

\section{Simulation Results}
\label{sec:experiments}
In this section, we present the simulation results to evaluate the performance of the proposed algorithm. We first describe the simulation settings and then analyze the results in detail.

\subsection{Setup}
All experiments are conducted in a simulated downlink OFDM network with four BSs placed at the corners of a $2000 \times 2000$ m square area. Each BS has a height of $30$ m and a maximum transmit power of $p_{\max} = 46$ dBm. The free-space pathloss model is adopted, where the channel attenuation from BS $n$ to UE $k$ is given by
$L^t_{k,n} = L_0 d_{k,n}^{\alpha}$,
in which $d_{k,n}$ denotes the distance between BS $n$ and UE $k$, $\alpha = 2.6$ is the pathloss exponent, and 
$L_0 = ( 4\pi f_c/ c)^{-2}$
is the channel power gain at a reference distance of $1$ m. The carrier frequency is set to $f_c = 5$ GHz, and the speed of light is $c = 10^8$ m/s. To simulate small-scale fading with temporal correlation, we use the first-order Gauss--Markov Rayleigh fading model~\cite{sharifi2021pomdp, he2024risk}:
\[
    h^t_{k,n,f} = \xi_{k,n} h^{t-1}_{k,n,f} + \delta^t_{k,n,f} \sqrt{1 - \xi_{k,n}^2},
\]
where $\xi_{k,n} \in [0,1]$ is the channel correlation coefficient, and $\delta^t_{k,n,f} \sim \mathcal{CN}(0,1)$ is the i.i.d.\ innovation term. The value of $\xi_{k,n}$ is determined by the maximum Doppler frequency, which in our experiments corresponds to a pedestrian user moving at $1$ m/s, yielding $\xi_{k,n} = 0.999$. The noise power spectral density is set to $-174$ dBm/Hz. Each training episode consists of $100$ steps, equivalent to $100$ frames or $1$ second in a 5G system. We evaluate performance under two network scales:
\begin{itemize}
    \item Small-scale setting: Each BS can serve up to $K_{\max} = 5$ users using a shared $20$ MHz bandwidth. With a subcarrier spacing of $720$ kHz, this results in $27$ subchannels per BS.
    \item {Large-scale setting:} Each BS can serve up to $K_{\max} = 20$ users using a shared $50$ MHz bandwidth. With a subchannel spacing of $120$ kHz, this results in $416$ subchannels per BS.
\end{itemize}
The above settings result in the size of the solution space of $5^{27}$ in the small-scale benchmark and $20^{416}$ in the large-scale benchmark, which is computationally infeasible for direct application of any naive RL method.

\begin{figure}[!b]
     \centering
     \begin{subfigure}{0.49\columnwidth}
         \centering
         \includegraphics[width=\textwidth]{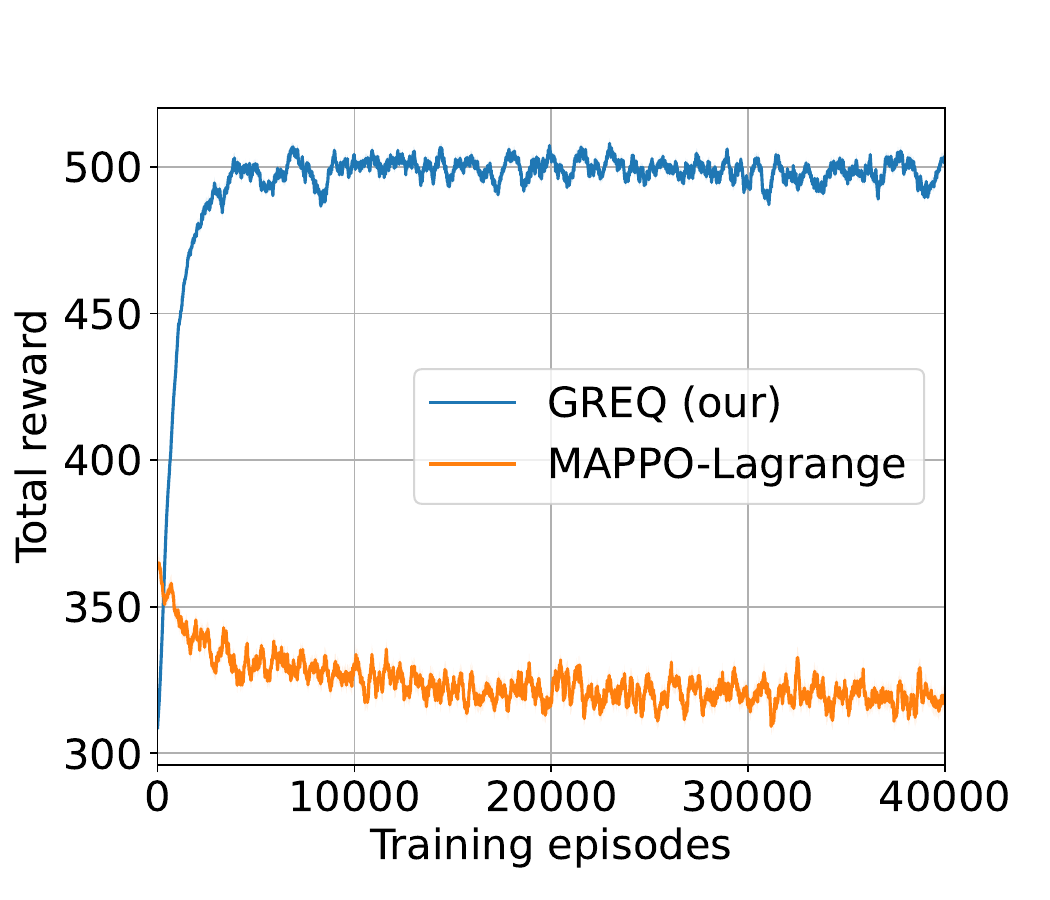}
         \caption{Reward}
     \end{subfigure}
     \hfill
     \begin{subfigure}{0.49\columnwidth}
         \centering
         \includegraphics[width=\textwidth]{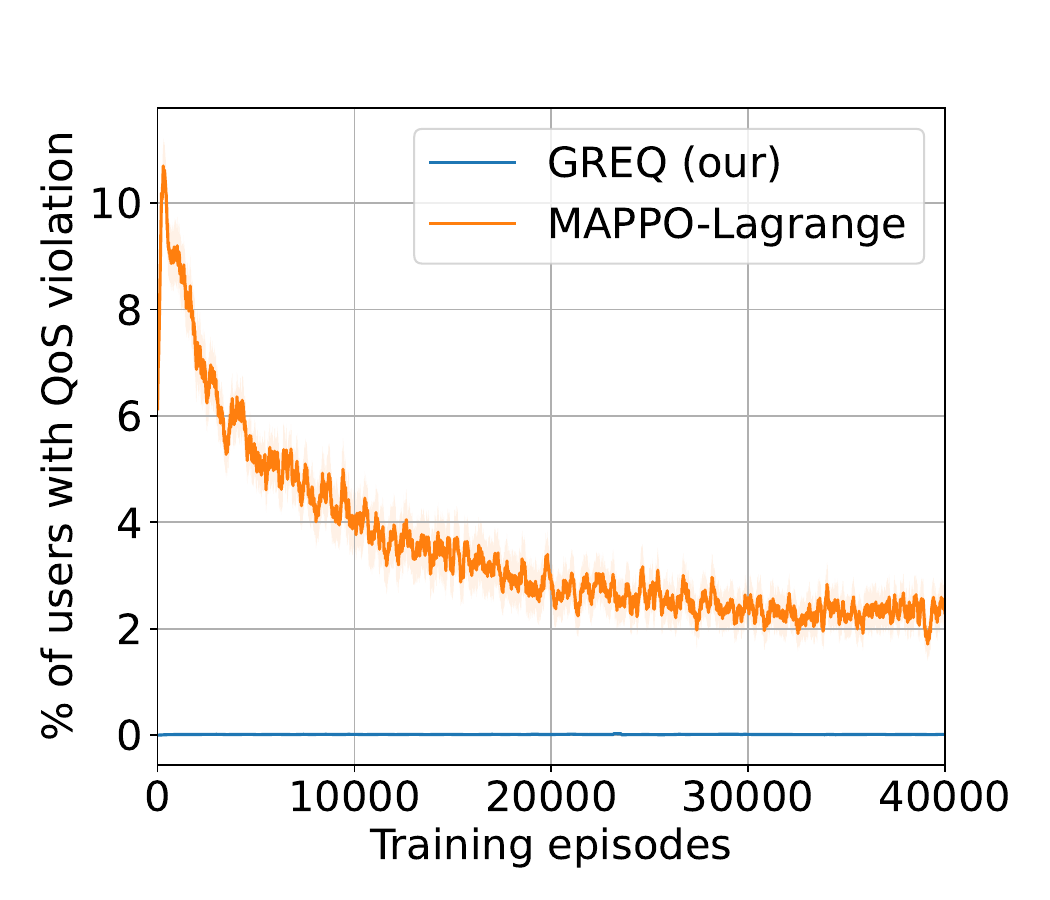}
         \caption{Constraint violation}
     \end{subfigure}
     \caption{Convergence trends of learning based algorithms.}
     \label{fig:learning_curve}
\end{figure}

We refer to the proposed algorithm as \textbf{GREQ}. Its performance is evaluated against the following baselines:
\begin{itemize}
    \item \textbf{DOPT}: A decentralized optimization approach to solve \eqref{prob:max_sum_rate_approx}. Successive convex approximation (SCA) technique with a penalty function is used to relax binary constraints of the allocation variables \cite{le2019energy,Fatemeh23}, followed by a rounding step to recover binary allocation solution. 
    \item \textbf{MAPPO-Lagrange}: A standard approach where multi-agent PPO framework equipped with Lagrangian relaxation to handle the QoS constraints in \eqref{prob:cmdp}. In this baseline, each BS is controlled by a single agent \cite{tang2020deep,yang2022distributed,naderializadeh2021resource} and all agents receive a common reward equal to the system sum-rate. A centralized value function is trained using the global state to stabilize learning. Lagrange multipliers are updated online to penalize QoS violations. All details of the learning environments are the same as those described in Subsection~\ref{subsec:env}, except that the spaces are the joint spaces of all $F$ subchannels. This is one of the most popular approaches for constrained MARL, but the large action space in our setting significantly challenges its scalability. More details of this framework can be found in \cite{gu2023safe}.
    \item \textbf{RRS}: A heuristic round-robin subchannel allocation procedure applied to every BS independently. The BS in turn assigns one user a strongest available subchannel, and repeats the procedure until all subchannels are used. This scheme requires only local CSI at each BS, however it also ignores potential inter-BS interference. 
    \item \textbf{RAND}: each BS randomly allocates subchannels to its users. This baseline provides a performance lower bound, highlighting the benefits of structure and learning in the other approaches.
\end{itemize}

\begin{figure*}
     \centering
     \begin{subfigure}{0.62\columnwidth}
         \centering
         \includegraphics[width=\textwidth]{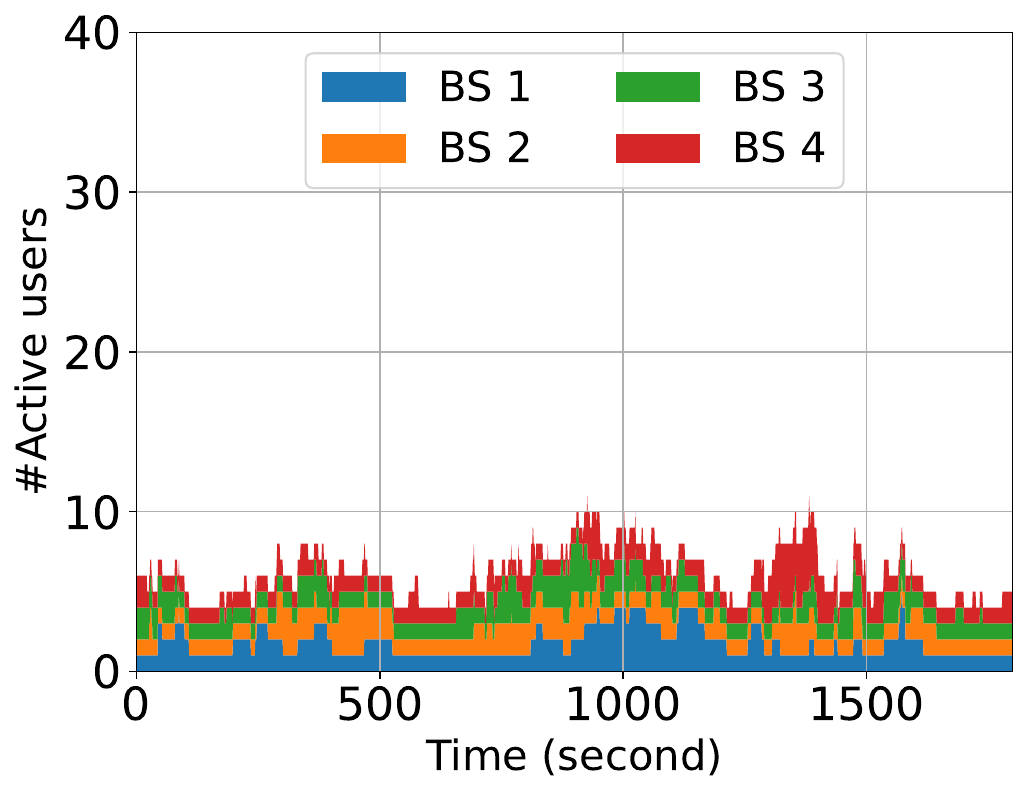}
         \caption{1 UE/minute/BS}
     \end{subfigure}
     \hfill
     \begin{subfigure}{0.62\columnwidth}
         \centering
         \includegraphics[width=\textwidth]{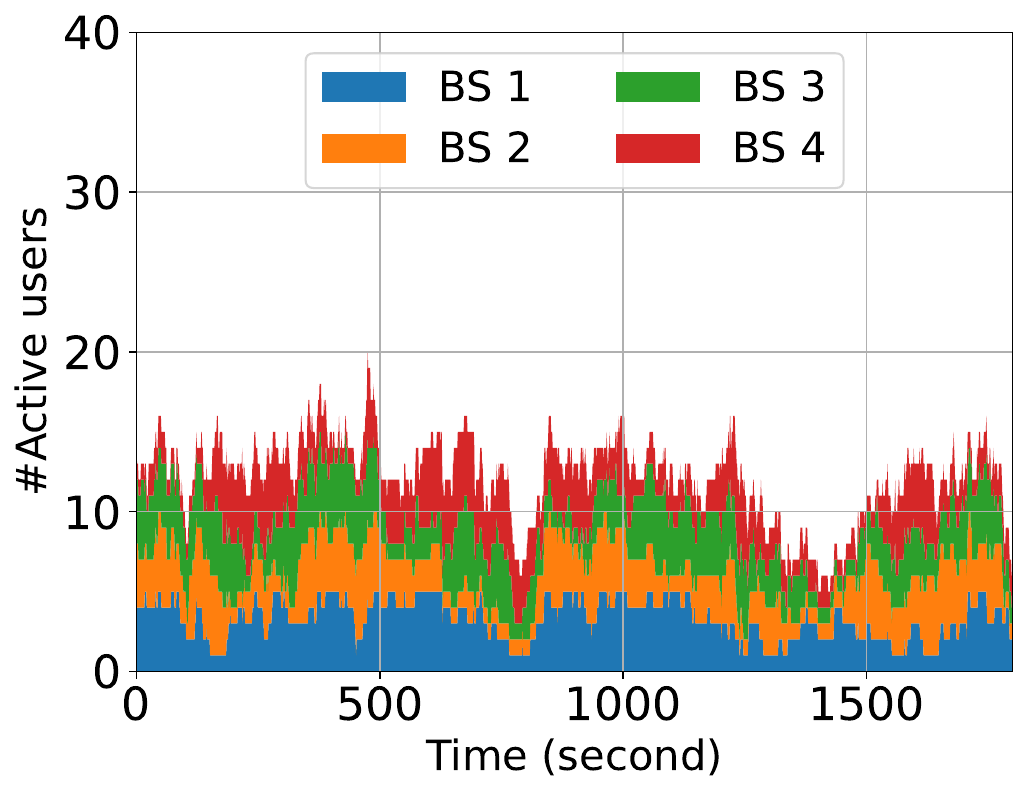}
         \caption{3 UEs/minute/BS}
     \end{subfigure}
     \hfill
     \begin{subfigure}{0.62\columnwidth}
         \centering
         \includegraphics[width=\textwidth]{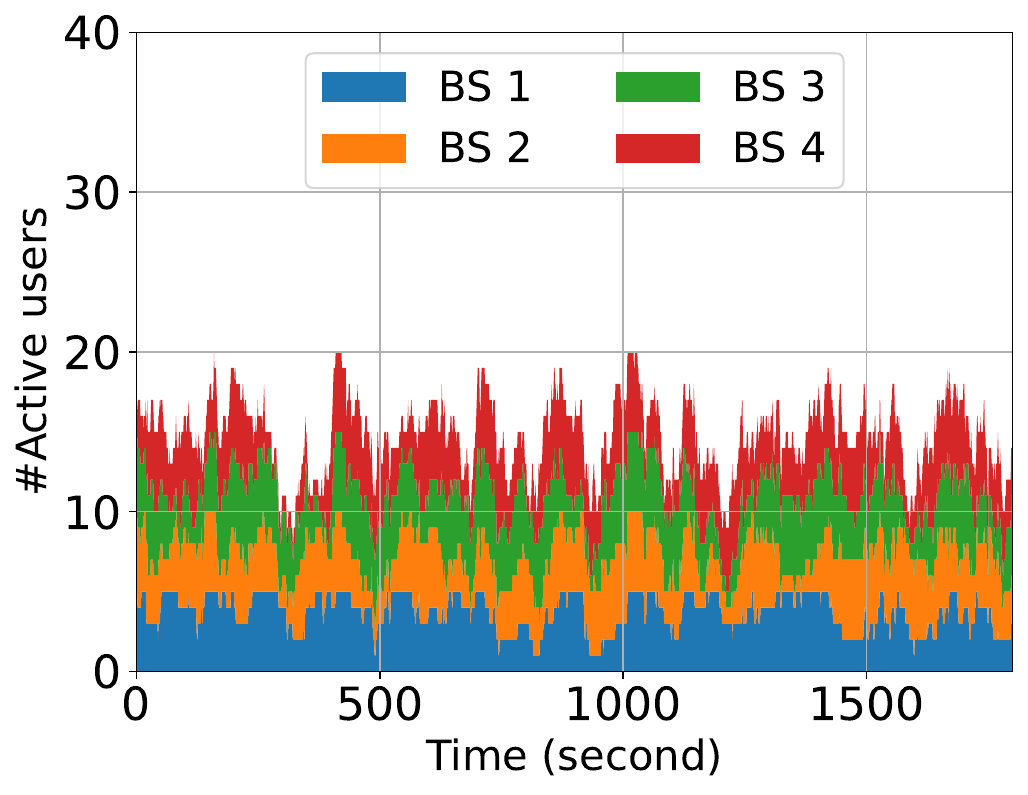}
         \caption{5 UEs/minute/BS}
     \end{subfigure}
    \caption{Number of active users at each BSs during the testing period. Each user stays in the network for average of 1 minute.}
     \label{fig:nb_users_small}
\end{figure*}

\begin{figure*}[h]
     \centering
     \begin{subfigure}{0.62\columnwidth}
         \centering
         \includegraphics[width=\textwidth]{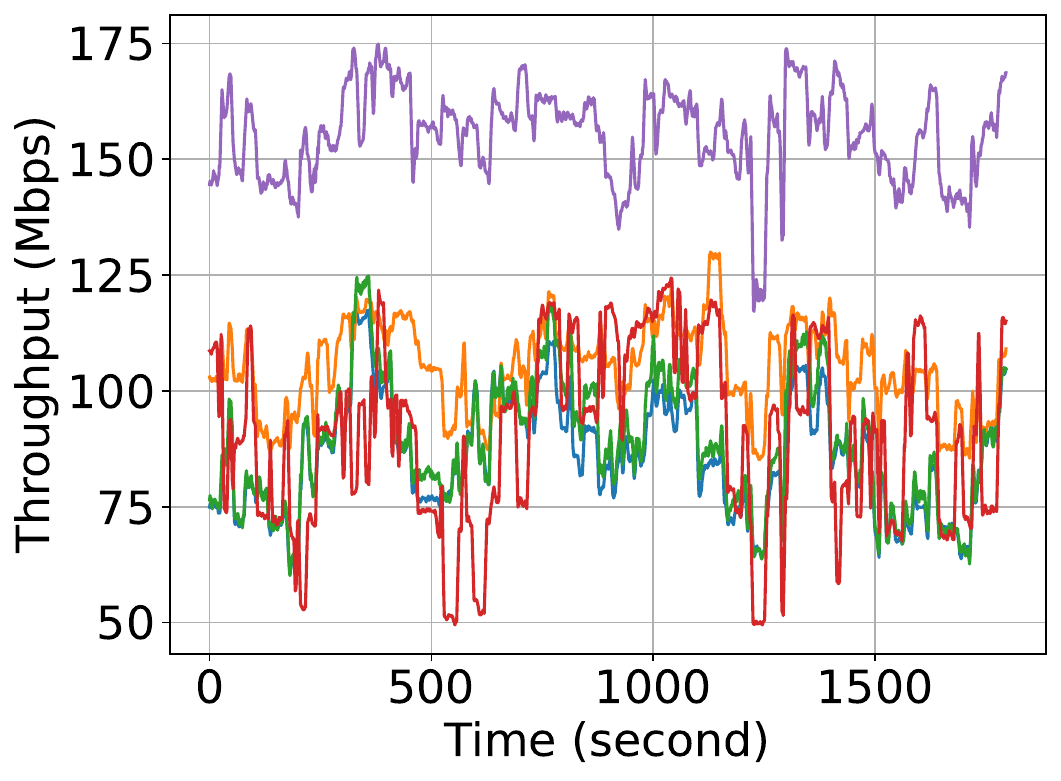}
         \caption{1 UE/minute/BS}
     \end{subfigure}
     \hfill
     \begin{subfigure}{0.62\columnwidth}
         \centering
         \includegraphics[width=\textwidth]{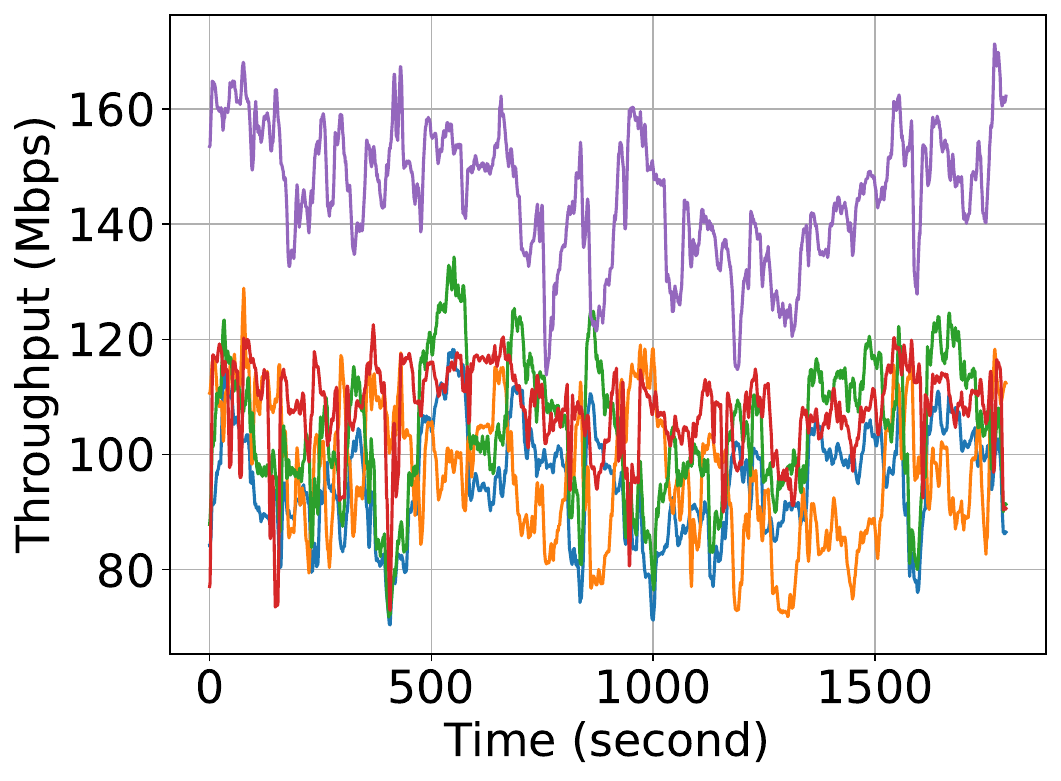}
         \caption{3 UEs/minute/BS}
     \end{subfigure}
     \hfill
     \begin{subfigure}{0.62\columnwidth}
         \centering
         \includegraphics[width=\textwidth]{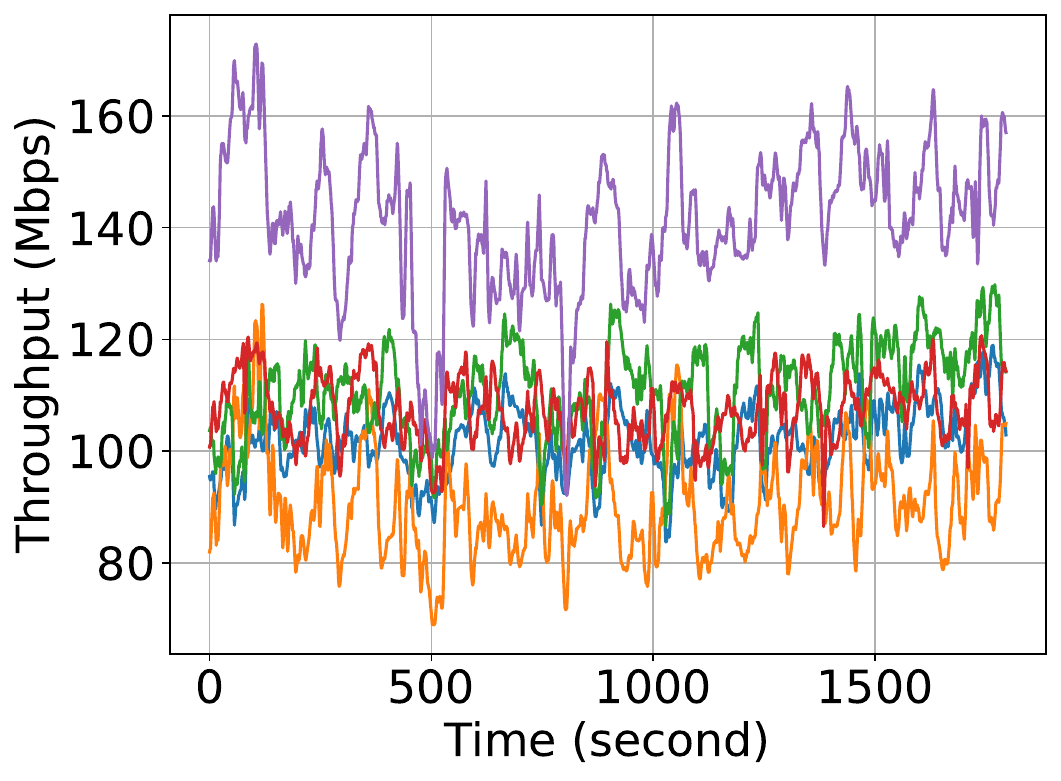}
         \caption{5 UEs/minute/BS}
     \end{subfigure}
     \hfill
     \begin{subfigure}{0.62\columnwidth}
         \centering
         \includegraphics[width=\textwidth]{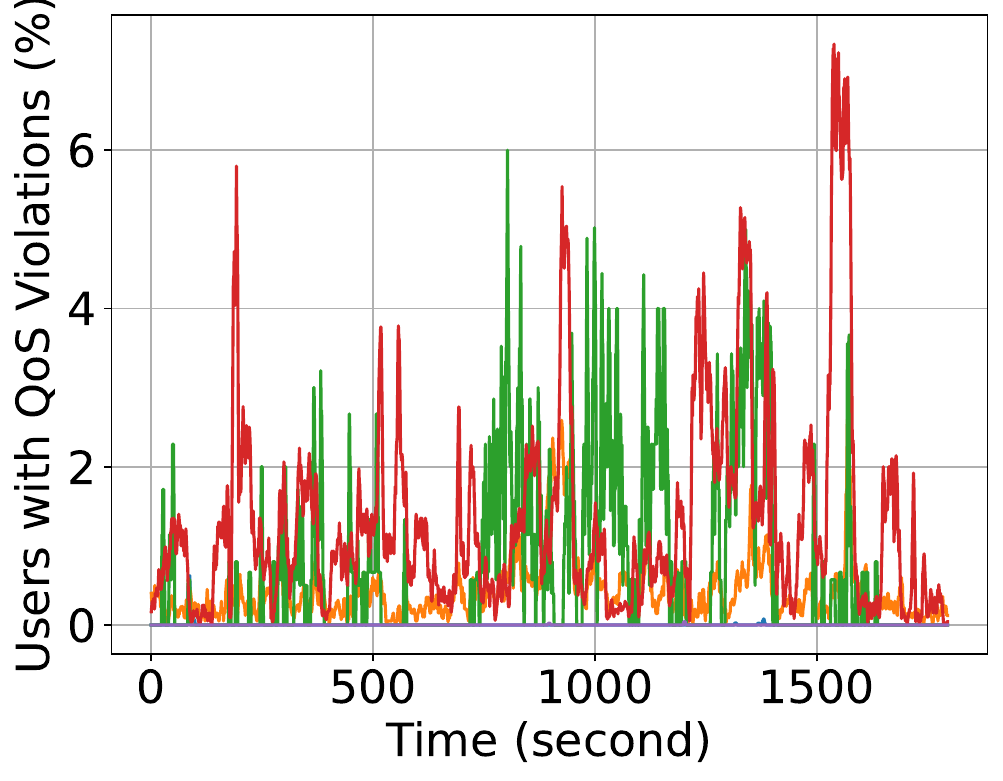}
         \caption{1 UE/minute/BS}
     \end{subfigure}
     \hfill
     \begin{subfigure}{0.62\columnwidth}
         \centering
         \includegraphics[width=\textwidth]{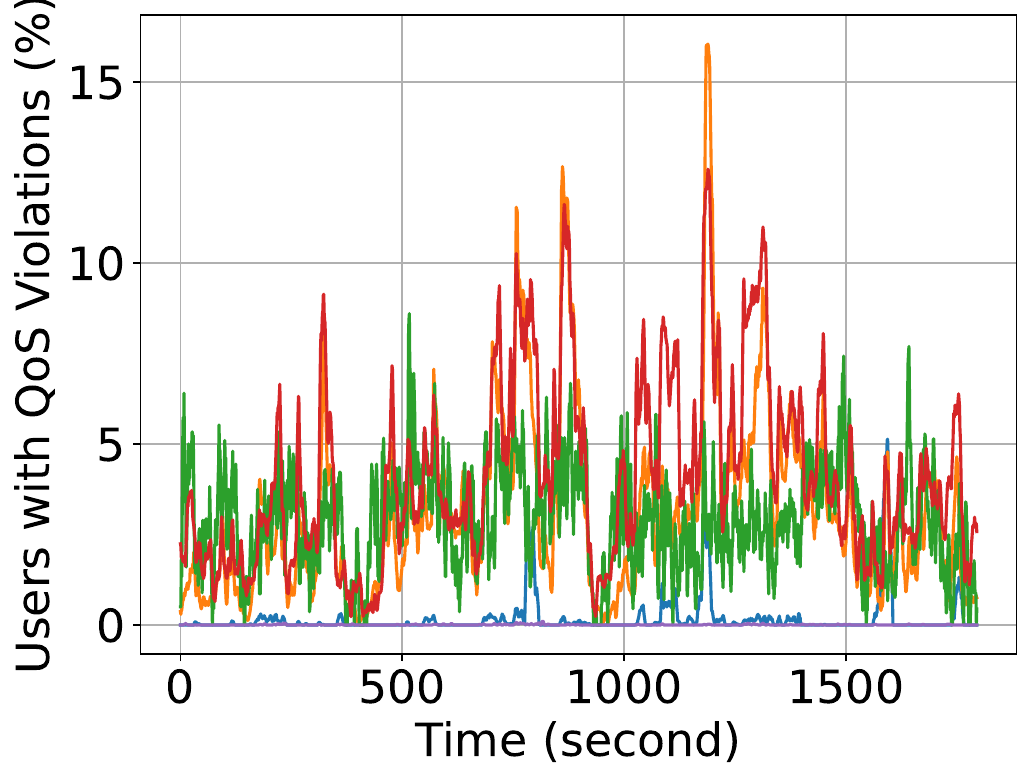}
         \caption{3 UEs/minute/BS}
     \end{subfigure}
     \hfill
     \begin{subfigure}{0.62\columnwidth}
         \centering
         \includegraphics[width=\textwidth]{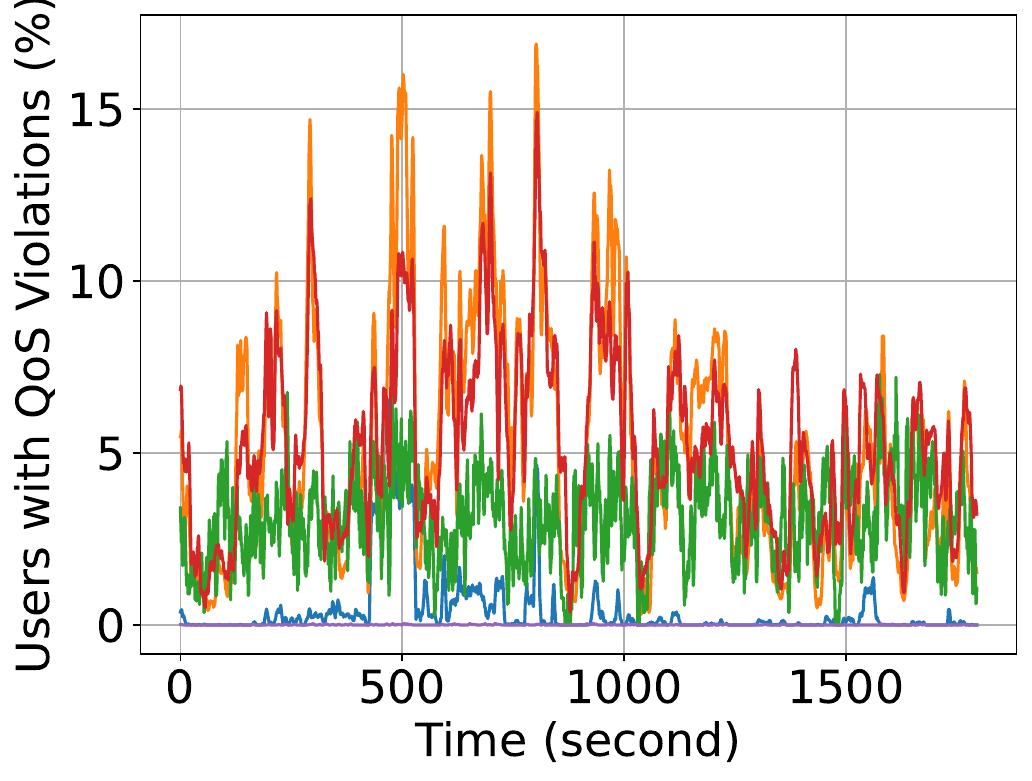}
         \caption{5 UEs/minute/BS}
     \end{subfigure}
     \hfill
     \begin{subfigure}{0.62\columnwidth}
         \centering
         \includegraphics[width=\textwidth]{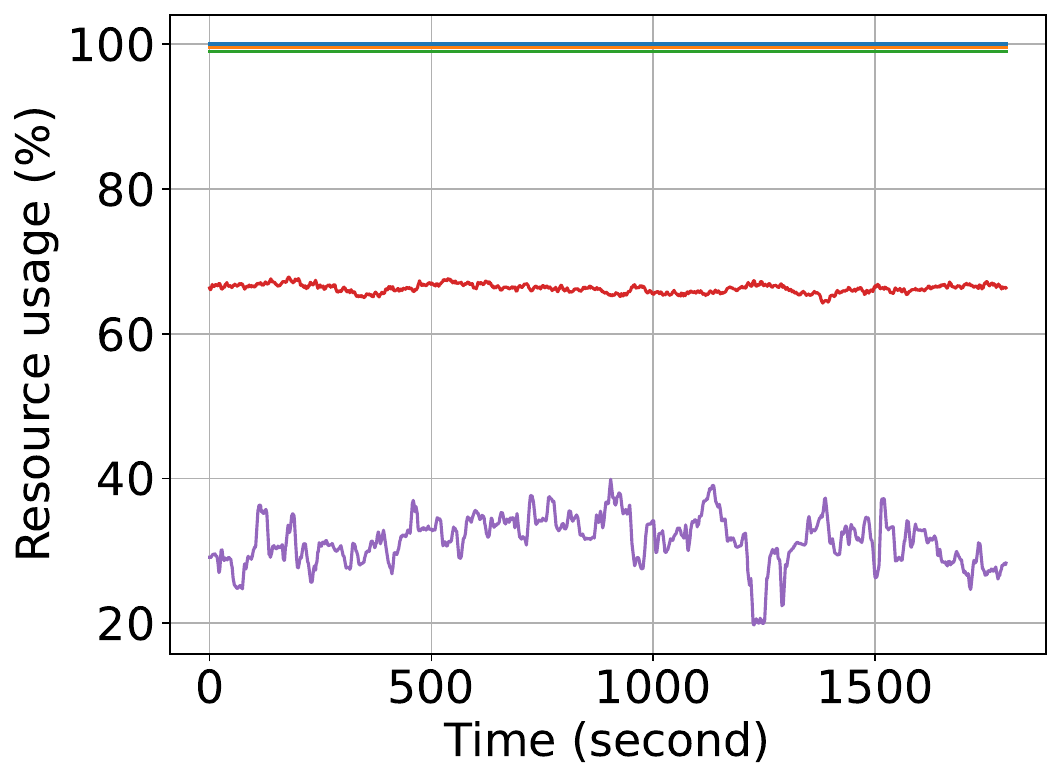}
         \caption{1 UE/minute/BS}
     \end{subfigure}
     \hfill
     \begin{subfigure}{0.62\columnwidth}
         \centering
         \includegraphics[width=\textwidth]{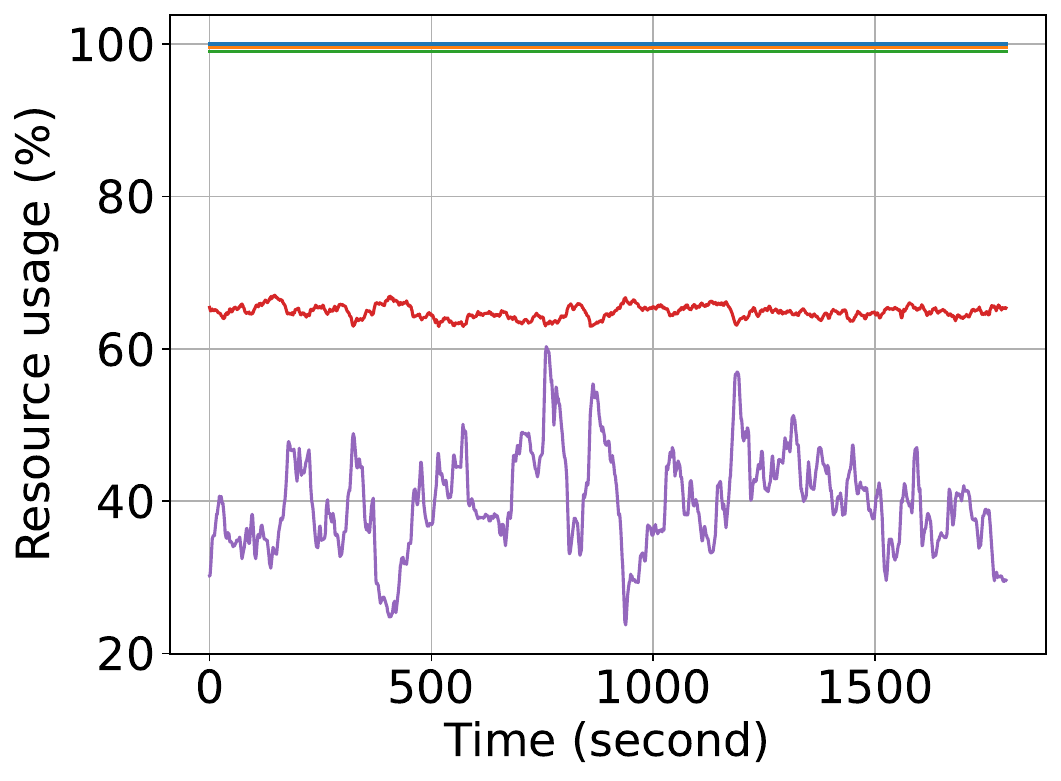}
         \caption{3 UEs/minute/BS}
     \end{subfigure}
     \hfill
     \begin{subfigure}{0.62\columnwidth}
         \centering
         \includegraphics[width=\textwidth]{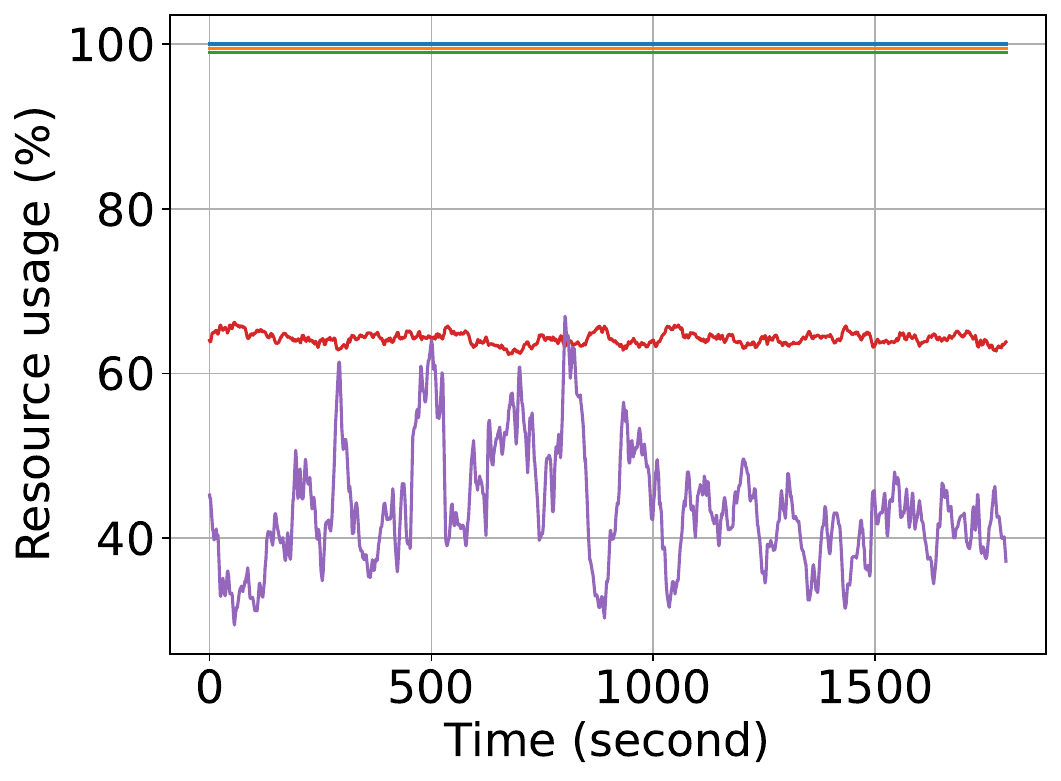}
         \caption{5 UEs/minute/BS}
     \end{subfigure}
     \hfill
     \begin{subfigure}{2\columnwidth}
         \centering
         \includegraphics[width=0.8\textwidth]{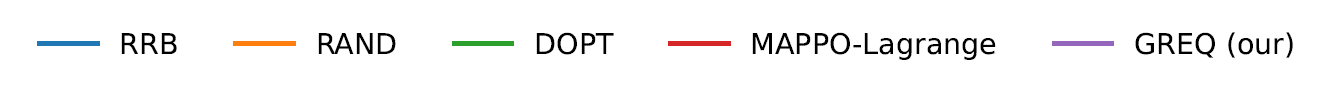}
     \end{subfigure}
     \caption{Testing results on small-scale benchmark.}
     \label{fig:results_small}
     \vspace{-0.5cm}
\end{figure*}

To expose the agents to the dynamics of the environment and prevent them from overfitting to specific user distributions or memorizing trivial allocation patterns, the set of active users, their QoS requirements, and their locations are re-generated at the start of each episode. In particular, the number of active users at each BS is drawn from a Poisson distribution with mean $\lambda = 0.5K_{\max}$, upper bounded by $K_{\max}$. Each user’s location is generated uniformly over the area and the user is associated with the closest BS. The QoS requirement of each user is drawn uniformly at random between 1 and 3 Mbps. This stochastic initialization introduces variability in both network topology and traffic load, encouraging the agents to learn generalizable allocation policies.  The MARL hyperparameters are tuned for stable convergence and reasonable performance. We use a learning rate of $5\times 10^{-5}$, a discount factor $\gamma = 0.9$, a batch size of 64, a replay buffer size of $10^5$ samples, and a target network update rate of $10^{-2}$. In both the agent utility networks and the QMIX mixing network, the number of hidden units in each layer is set to 128.

\subsection{Results on Small-scale Setting}
Fig.~\ref{fig:learning_curve} shows the training performance of the two learning-based methods on the small-scale setting: the proposed GREQ algorithm (blue curves) and the MAPPO-Lagrange baseline (orange curves). Subfigure (a) reports the total episode reward, while subfigure (b) presents the average percentage of users experiencing QoS violations during training. GREQ achieves a rapid increase in episode reward while maintaining a consistent zero violation rate throughout training. In contrast, MAPPO-Lagrange attains significantly lower rewards and suffers from persistently higher violation rates. Notably, as MAPPO-Lagrange reduces its constraint violation rate over time, its episode reward also declines. This behavior arises because the initial policies of MAPPO-Lagrange start with high violation rates. During training, the increasing Lagrange multiplier shifts the optimization focus toward satisfying the QoS constraints, often at the expense of throughput. However, even with this trade-off, MAPPO-Lagrange fails to eliminate violations entirely, unlike GREQ. The results in Fig.~\ref{fig:learning_curve}b highlight the strong effectiveness of the constraint-handling mechanism of GREQ. By offloading QoS satisfaction to a deterministic greedy procedure in the first phase, the second phase only needs to solve unconstrained subproblems with much smaller action spaces, enabling faster reward improvement and quicker convergence.

Next, we test the trained policy in a dynamic environment mimicking real scenarios. In particular, new users arrive according to a Poisson process with mean arrival rates of 1, 3, and 5 users per minute, where higher arrival rates naturally result in a larger and more dynamic active user set. For this small-scale test case, the duration of each user’s stay in the system is modeled using an exponential distribution with a mean (scale parameter) of 1 minute, ensuring a continuous turnover of active users. During their stay, users may change location following a random-walk model. To capture this mobility, each user’s location is updated every second with probability 0.5 and moving at a constant speed of 1 m/s. This setup allows us to examine the algorithms’ robustness under varying user densities, dynamic arrivals, and mobility patterns. All methods are evaluated over a simulation horizon of 1800 seconds (30 minutes), enabling the collection of statistically meaningful performance data. Figure~\ref{fig:nb_users_small} illustrates the temporal evolution of the number of active users in the system, with subfigures (a), (b), and (c) corresponding to scenarios where 1, 3, and 5 users arrive per minute on average, respectively.

\begin{figure}
     \centering
     \begin{subfigure}{0.98\columnwidth}
         \centering
         \includegraphics[width=\textwidth]{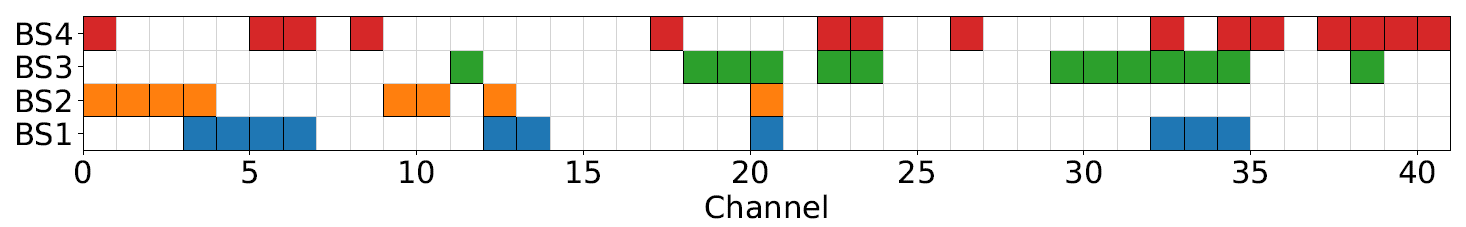}
         \caption{1 UE/minute/BS}
     \end{subfigure}
     \hfill
     \begin{subfigure}{0.98\columnwidth}
         \centering
         \includegraphics[width=\textwidth]{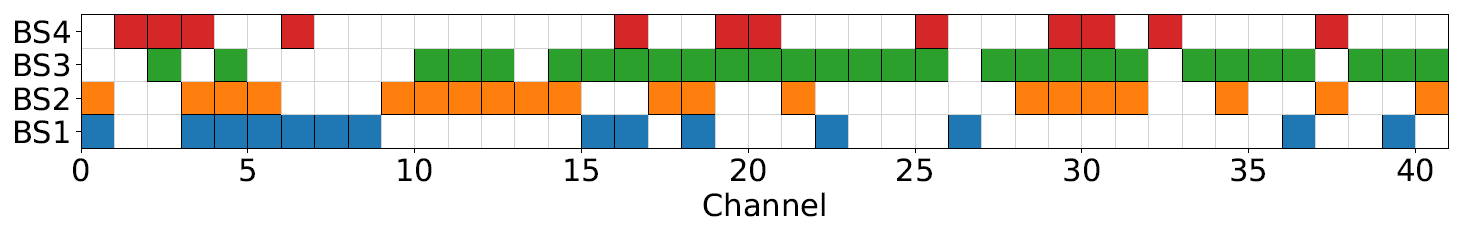}
         \caption{3 UEs/minute/BS}
     \end{subfigure}
     \hfill
     \begin{subfigure}{0.98\columnwidth}
         \centering
         \includegraphics[width=\textwidth]{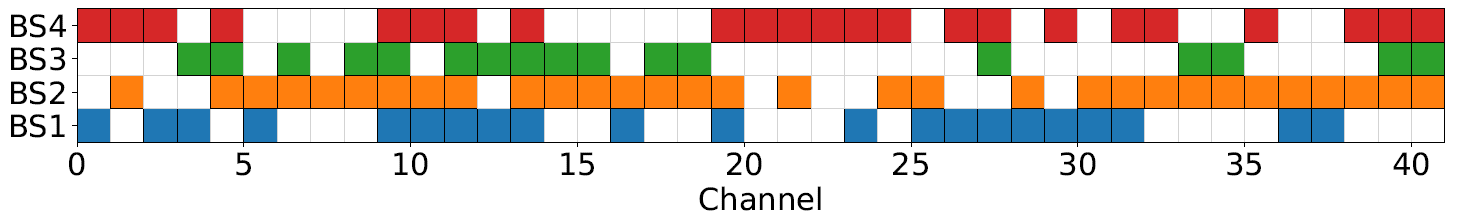}
         \caption{5 UEs/minute/BS}
     \end{subfigure}
     \caption{Example of subchannel assignment produced by the proposed framework on a small test instance. Colored squares represent subchannels activated by the corresponding BS, while uncolored squares indicate deactivated subchannels.}
     \label{fig:nb_user_small_assignment}
\end{figure}
\begin{figure*}[h]
     \centering
     \begin{subfigure}{0.62\columnwidth}
         \centering
         \includegraphics[width=\textwidth]{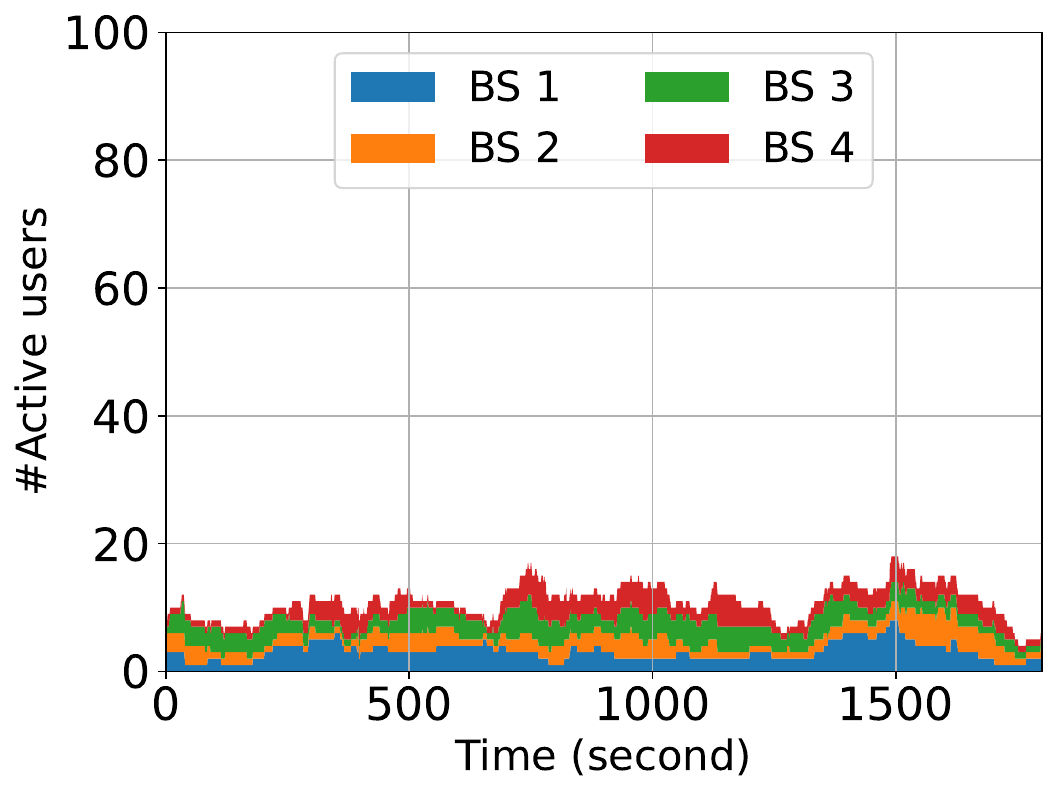}
         \caption{1 UE/minute/BS}
     \end{subfigure}
     \hfill
     \begin{subfigure}{0.62\columnwidth}
         \centering
         \includegraphics[width=\textwidth]{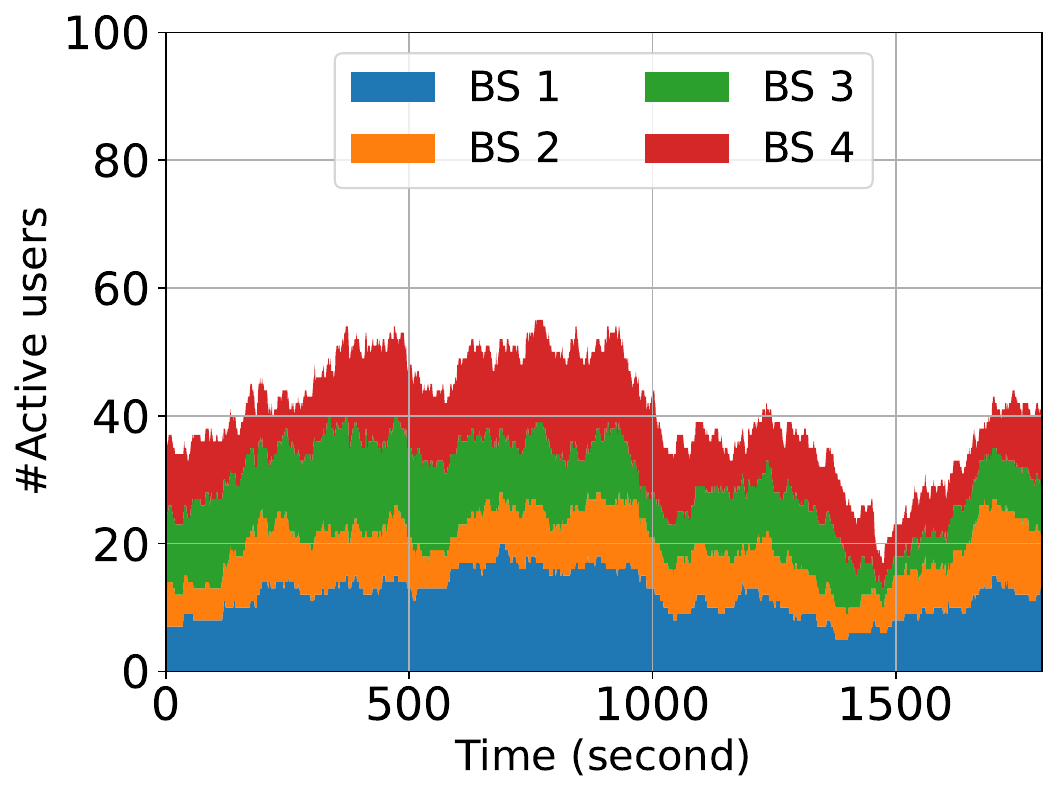}
         \caption{3 UEs/minute/BS}
     \end{subfigure}
     \hfill
     \begin{subfigure}{0.62\columnwidth}
         \centering
         \includegraphics[width=\textwidth]{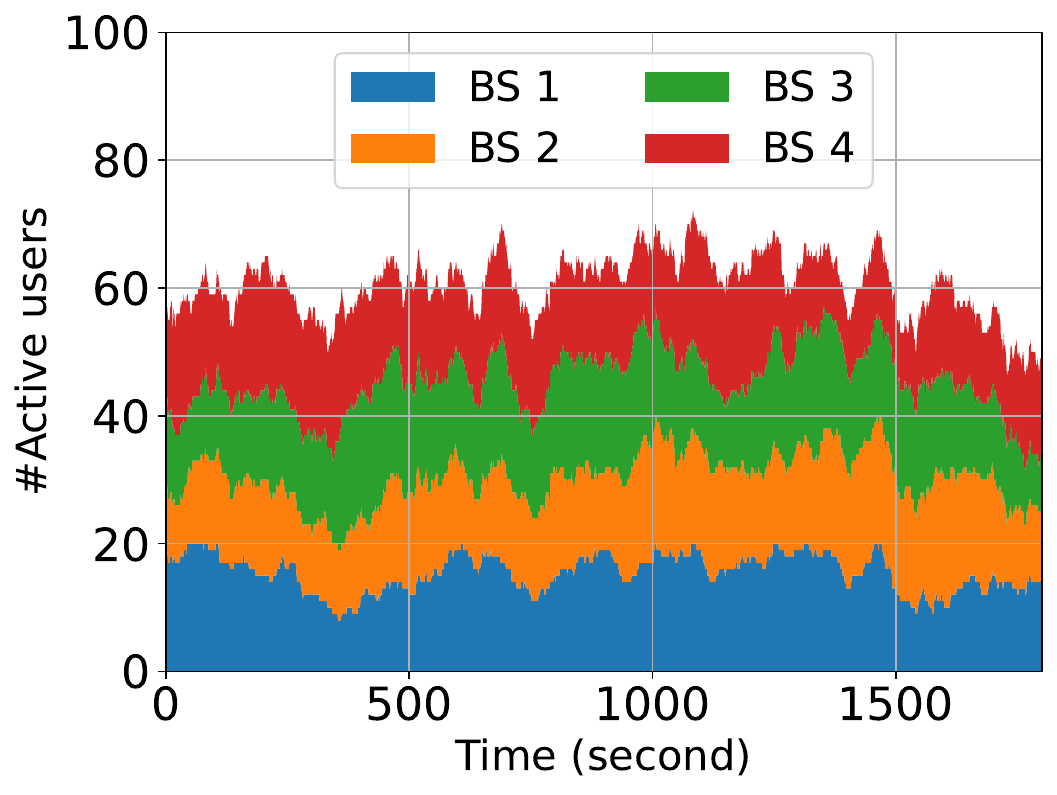}
         \caption{5 UEs/minute/BS}
     \end{subfigure}
    \caption{Number of active users at each BSs during the testing period. Each user stays in the network for average of 3 minutes.}
     \label{fig:nb_users_large}
     \vspace{-0.5cm}
\end{figure*}

Fig.~\ref{fig:results_small} presents the testing results for all evaluated algorithms. Vertically, the first, second, and third columns correspond to user arrival rates of 1, 3, and 5 users per minute per BS, respectively. Horizontally, the top row reports overall system throughput, the middle row shows the percentage of users with QoS violations, and the bottom row depicts resource utilization (the portion of subchannels activated during the testing period). Across all scenarios, GREQ achieves the highest throughput and maintains very low (at zero most of the time) QoS violations, while activating a smallest number of subchannels (to efficiently mitigate inter-cell interference). This performance is driven by GREQ's design in which the MARL phase learns to selectively activate or deactivate subchannels to mitigate inter-cell interference while maximizing sum-rate. By contrast, MAPPO-Lagrange performs poorly, with lower throughput and higher QoS violation rates compared to GREQ. This result is consistent with its training curves in Fig.~\ref{fig:learning_curve}. The remaining three methods, including RRS, RAND, and DOPT, show similar performances, with system throughput approximately 30\% lower than GREQ, despite using 100\% of the available system resources. These methods fail to meet the QoS requirements of every user. However, RRS significantly outperforms RAND and DOPT in terms of QoS violation. The relatively low violation rate of RRB stems from its round-robin scheduling, which ensures that each user takes turns selecting high-quality subchannels, allowing most of the users to obtain enough resources to meet their QoS requirements. For DOPT, each BS attempts to maximize its own performance independently, without any coordination mechanism to mitigate interference. As a result, all subchannels are activated selfishly, resulting in more inter-cell interference and consequently higher probability of QoS violation.

\begin{figure*}[h]
     \centering
     \begin{subfigure}{0.62\columnwidth}
         \centering
         \includegraphics[width=\textwidth]{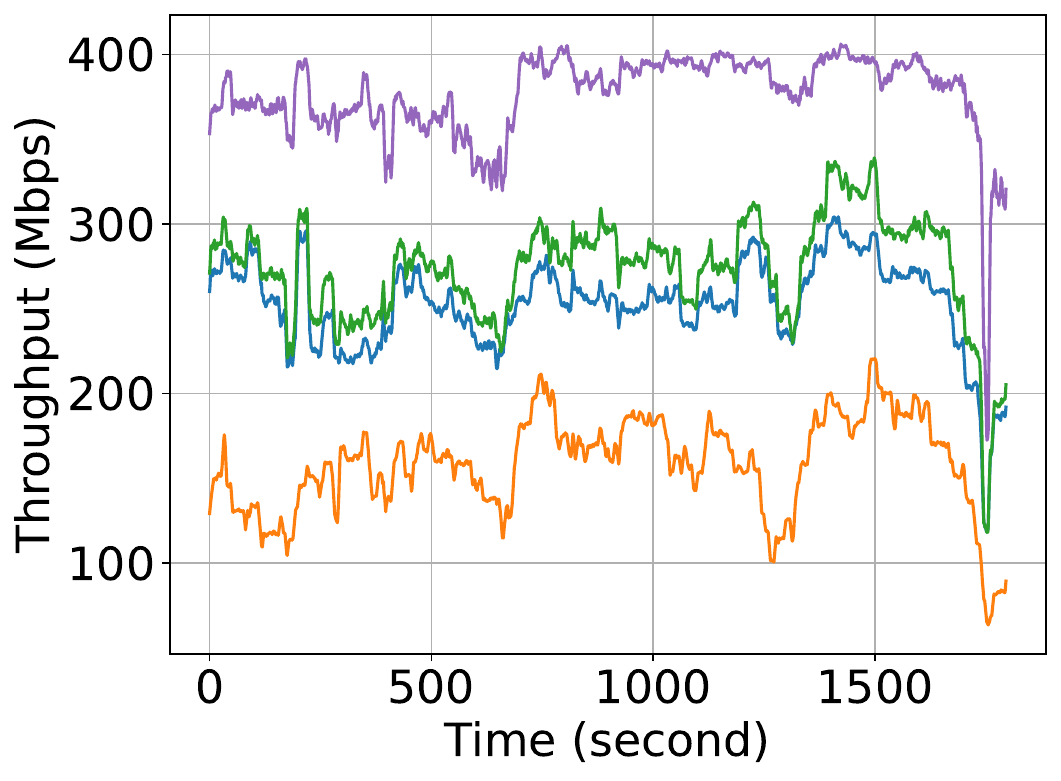}
         \caption{1 UE/minute/BS}
     \end{subfigure}
     \hfill
     \begin{subfigure}{0.62\columnwidth}
         \centering
         \includegraphics[width=\textwidth]{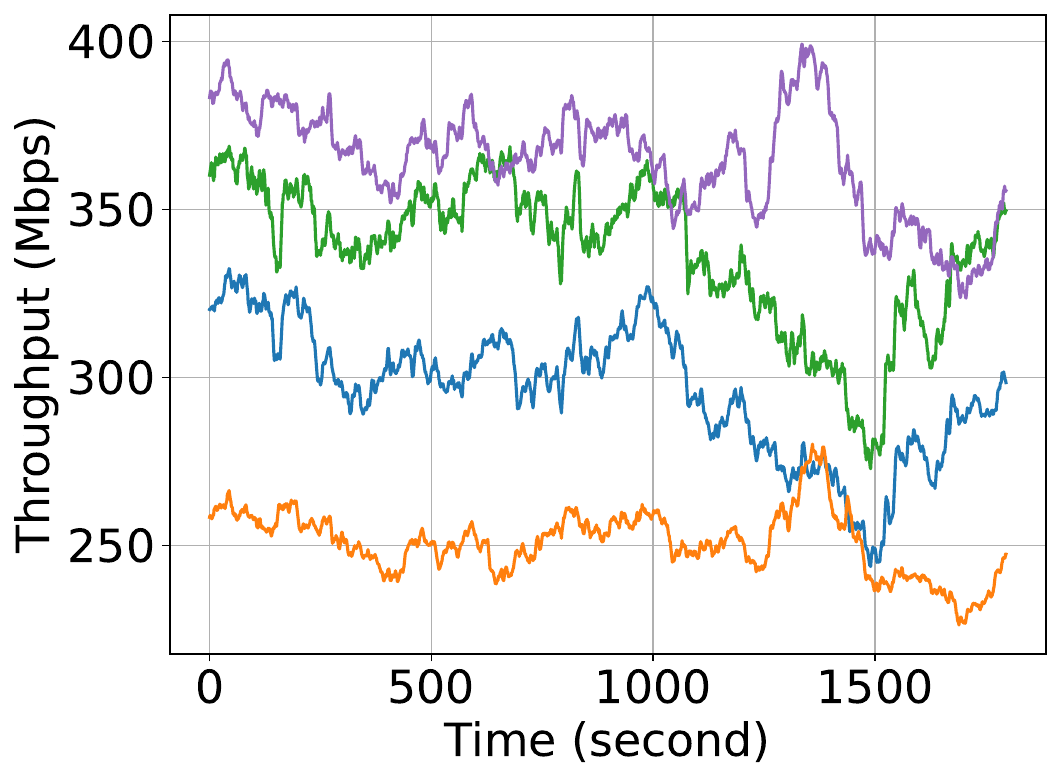}
         \caption{3 UEs/minute/BS}
     \end{subfigure}
     \hfill
     \begin{subfigure}{0.62\columnwidth}
         \centering
         \includegraphics[width=\textwidth]{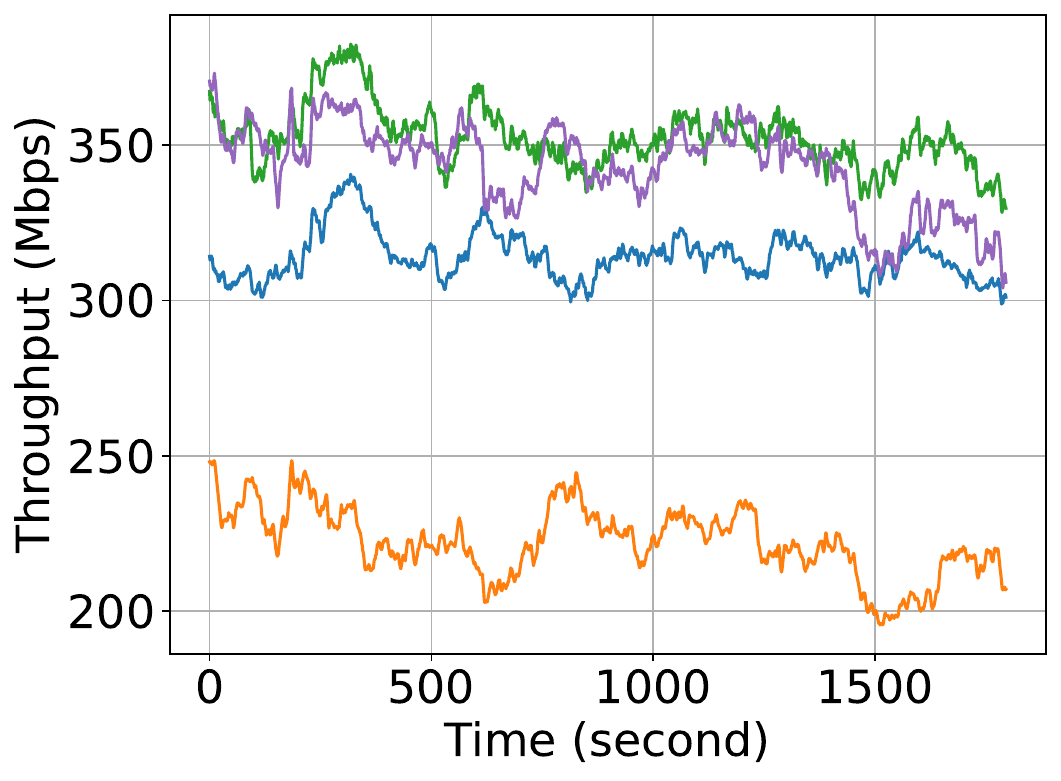}
         \caption{5 UEs/minute/BS}
     \end{subfigure}
     \hfill
     \begin{subfigure}{0.62\columnwidth}
         \centering
         \includegraphics[width=\textwidth]{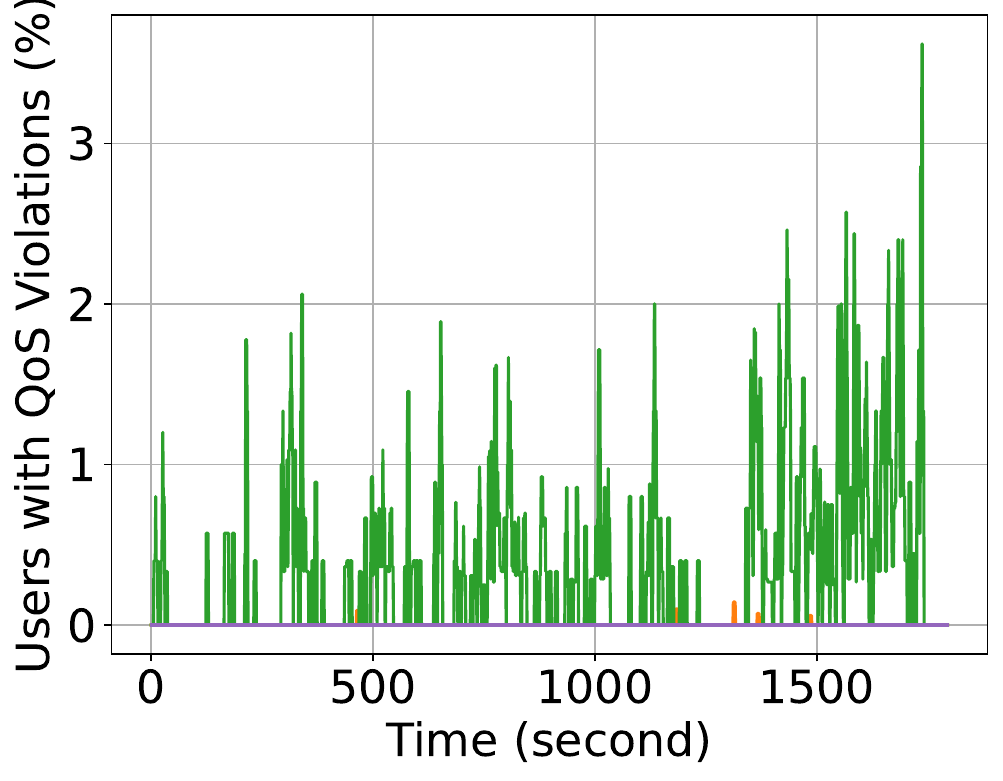}
         \caption{1 UE/minute/BS}
     \end{subfigure}
     \hfill
     \begin{subfigure}{0.62\columnwidth}
         \centering
         \includegraphics[width=\textwidth]{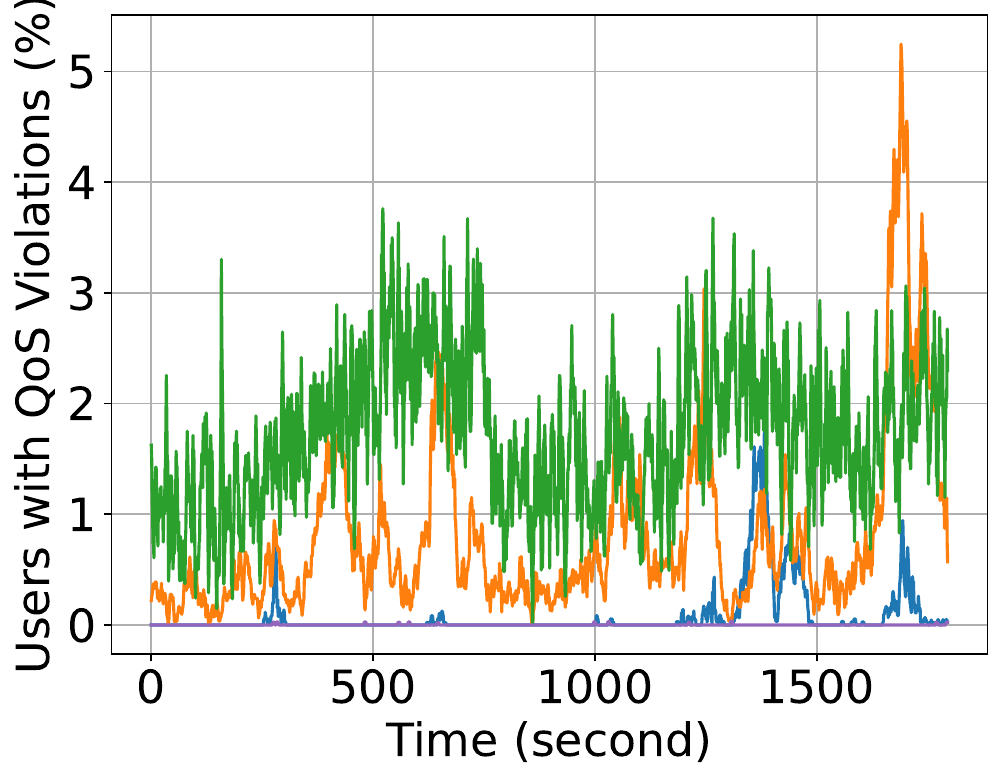}
         \caption{3 UEs/minute/BS}
     \end{subfigure}
     \hfill
     \begin{subfigure}{0.62\columnwidth}
         \centering
         \includegraphics[width=\textwidth]{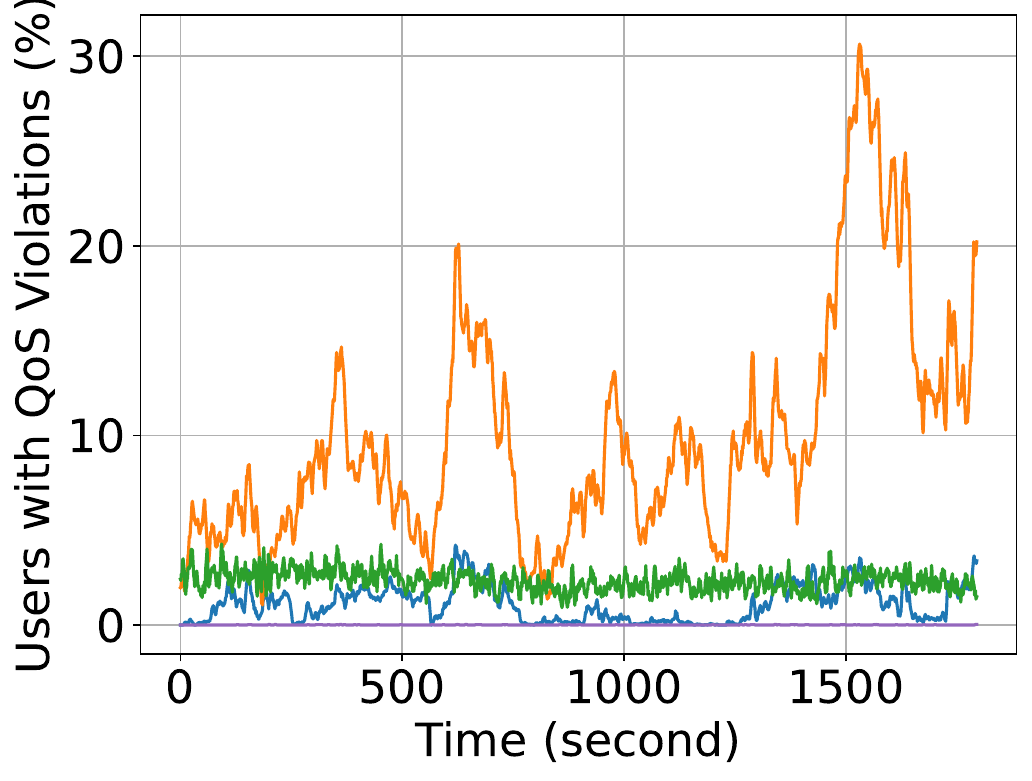}
         \caption{5 UEs/minute/BS}
     \end{subfigure}
     \hfill
     \begin{subfigure}{0.62\columnwidth}
         \centering
         \includegraphics[width=\textwidth]{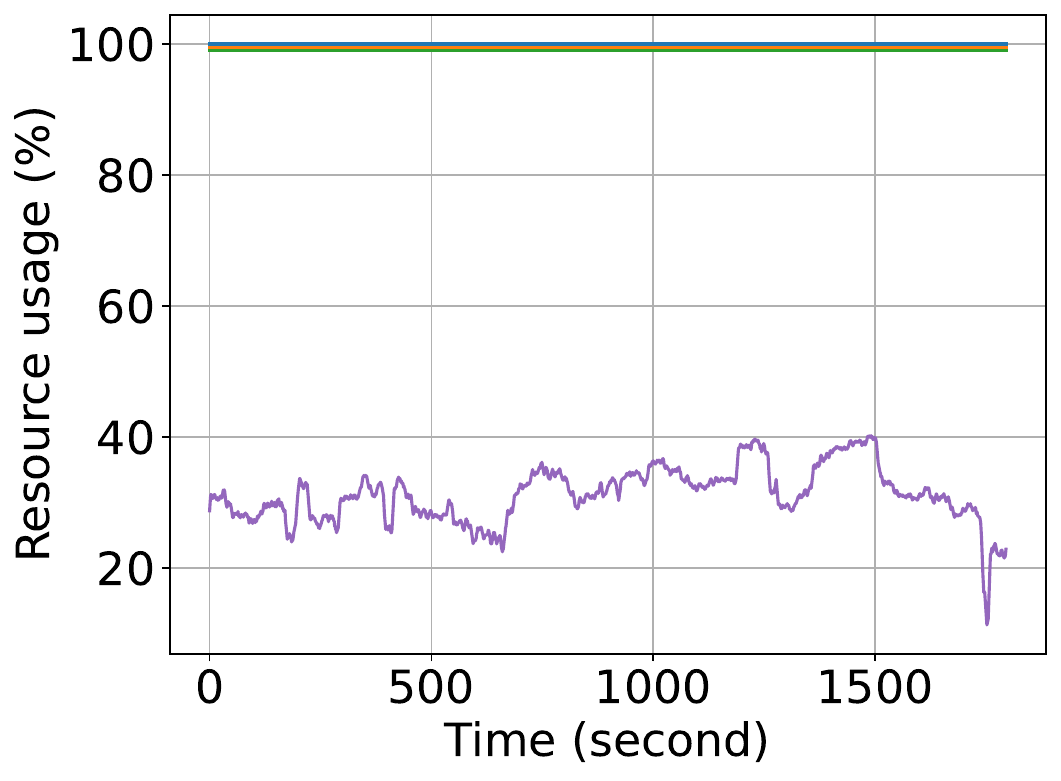}
         \caption{1 UE/minute/BS}
     \end{subfigure}
     \hfill
     \begin{subfigure}{0.62\columnwidth}
         \centering
         \includegraphics[width=\textwidth]{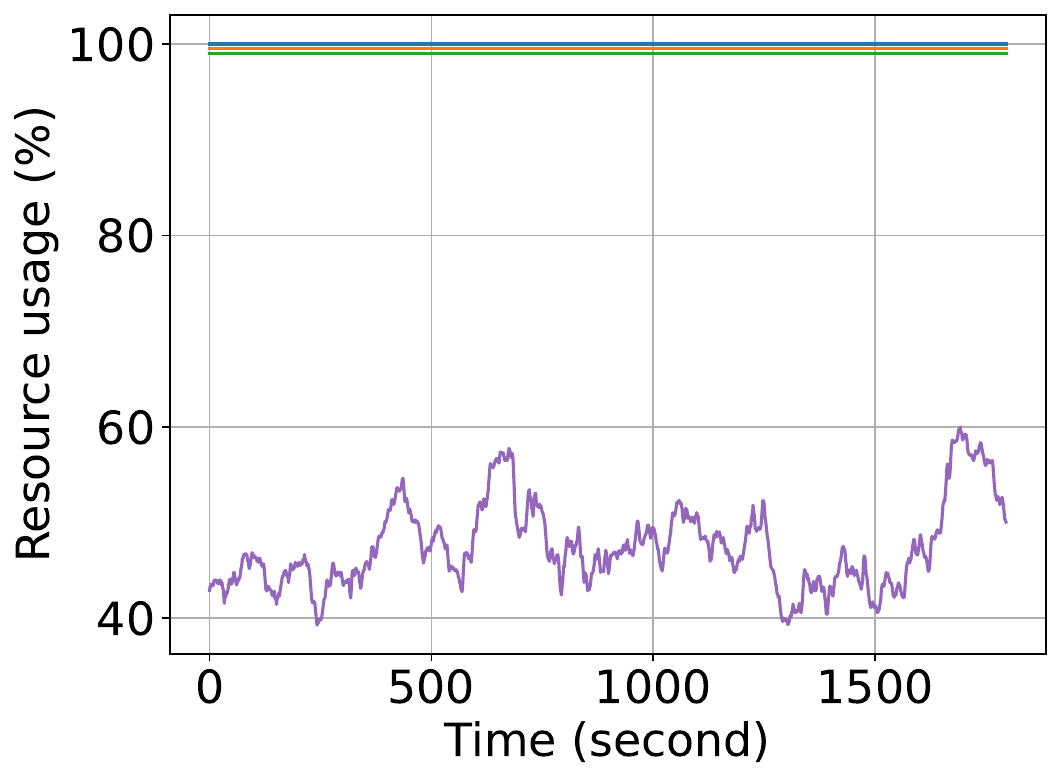}
         \caption{3 UEs/minute/BS}
     \end{subfigure}
     \hfill
     \begin{subfigure}{0.62\columnwidth}
         \centering
         \includegraphics[width=\textwidth]{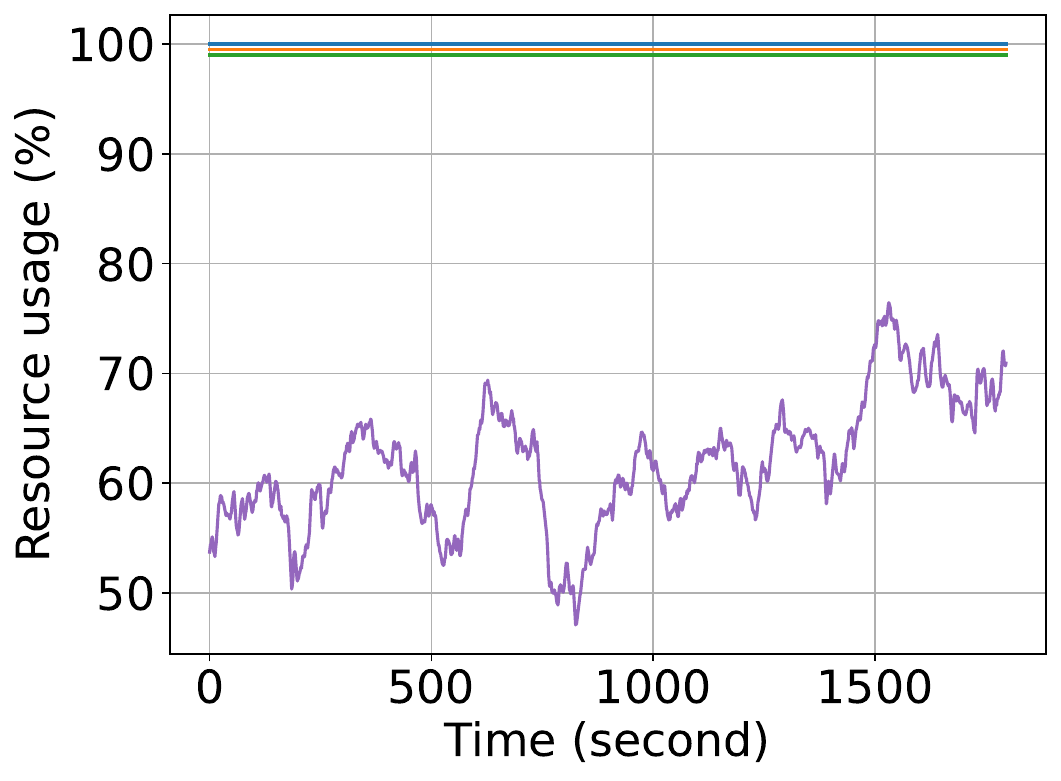}
         \caption{5 UEs/minute/BS}
     \end{subfigure}
     \hfill
     \begin{subfigure}{2\columnwidth}
         \centering
         \includegraphics[width=0.6\textwidth]{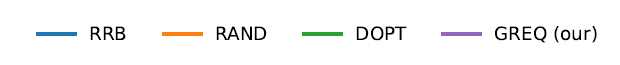}
     \end{subfigure}     
     \vspace{-0.3cm}
     \caption{Testing results on large-scale benchmark.}
     \label{fig:results_large}
\end{figure*}

In Fig.~\ref{fig:nb_user_small_assignment}, we present sample resource allocation snapshots at a single time step generated by the proposed GREQ algorithm under varying user arrival rates. Each subfigure illustrates how each BS selectively activates its subchannels. The results demonstrate that the agents have effectively learned to deactivate certain subchannels to mitigate inter-BS interference, rather than activating all available resources selfishly. It can also be seen that as the user arrival rate increases, more subchannels are activated to accommodate the higher demand and satisfy the QoS requirements. This behavior reflects the agents’ ability to dynamically balance the trade-off between maximizing system throughput and minimizing interference. It is worth noting that since the transmit power per subchannel is fixed and uniformly allocated, activating fewer subchannels also reduces total power consumption. Therefore, compared to baseline methods that activate all subchannels, GREQ’s selective activation not only mitigates interference but also achieves more efficient power usage, which is an important advantage in wireless systems.

\subsection{Results on Large-scale Setting}
In this subsection, we present the testing results on the large-scale setting, where each BS can serve up to 20 users using a 50 MHz bandwidth divided into 416 subchannels. The simulation environment retains the same configurations as the small-scale setting in terms of user mobility and arrival process. However, the average active duration of each user is increased to three minutes, resulting in more simultaneous active users in the system. This setup better reflects realistic high-load conditions and allows us to evaluate the scalability and robustness of the proposed algorithm in more demanding scenarios. Figure~\ref{fig:nb_users_large} plots the evolution of the number of users in the system, with subfigures (a), (b), and (c) corresponding to scenarios where 1, 3, and 5 users arrive per minute on average, respectively.

Fig.~\ref{fig:results_large} presents the testing results of four methods, including RRS, RAND, DOPT, and the proposed GREQ. The MAPPO-Lagrange baseline is excluded from this experiment due to the prohibitive size of the action space. The observed trends are consistent with the small-scale results. Specifically, GREQ achieves the highest system throughput in most scenarios, maintains the lowest QoS violation (nearly zero across all cases), and utilizes the least resources. Remarkably, GREQ activates only 20–70\% of the available subchannels, depending on the network load. By contrast, the random allocation baseline RAND performs poorly, with substantial lower throughput and higher QoS violations, especially as user arrival rates increase. Both RRS and DOPT maintain QoS violations at a moderate level of approximately 3\%. However, these two baselines activate all available subchannels regardless of network load, leading to excessive interference and inefficient power usage. This results again confirm the advantage of GREQ’s intelligent subchannel activation and assignment.

Overall, the proposed solution GREQ demonstrates impressive performance, consistently achieving the highest system throughput in most scenarios while utilizing only 20–80\% of the available resources depending on the network load. Most importantly, thanks to the first greedy-assignment phase, GREQ is remarkably effective in satisfying users’ QoS requirements, maintaining almost zero constraint violation in all tested cases. In contrast, the naive MARL solution fails to learn in large-scale problem sizes due to prohibited action space cardinality. This finding reinforces the motivation for our hybrid two-phase design and underscores the practical value of integrating domain knowledge with learning-based methods in wireless resource management.

\section{Conclusions and Discussions} 
\label{sec:conclusion}
This paper proposed a hybrid model-learning framework for distributed radio resource management in large-scale OFDM-based wireless systems. Although MARL offers a powerful paradigm for decentralized control, our study showed that directly applying it to the full-scale resource allocation problem is computationally infeasible due to the enormous joint action space and dynamics of the learning environment.
To address these challenges, we first allocate a minimum resource to satisfy the QoS requirements using a model-based greedy algorithm. In the second phase, a scalable multiagent system allocates the remaining resources to further enhance the system throughput. To train the multiagent system for the second phase, we exploit the orthogonality property of OFDM transmissions to decompose the original problem into subproblems which can be trained in parallel, thereby overcome the issue of action space explosion.
Together, these techniques enable scalable and effective distributed learning without compromising performance. This work highlights the importance of domain-aware hybrid model-learning designs in wireless systems, where model-based and learning-based methods can be integrated to overcome the limitations of pure MARL approaches.

\emph{Discussions and Perspectives:} The proposed hybrid scalable learning framework has shown impressive performance in both constraint satisfaction and scalability, however, the results were performed in a simulation environment. Towards efficient deployment, the trained model should be fine-tuned with real data sets obtained via a measurement campaign. This fine-tuning is essential for the model to adapt to the real inherent data features that are not generated by the simulation environment. Another important aspect is potential model mismatches when testing the trained model in a field test, which is due to the large space of the real propagation environment that the simulation cannot sufficiently represent. In this case, multi-modal or online learning method should be deployed to adjust the policies according to the actual environment in real-time. These can be implemented in the Ran Intelligent Controller (RIC) module in the Open RAN architecture. For large networks with many BSs sharing the spectrum, we may employ a divide-and-conquer strategy to form clusters of BSs and apply the proposed method to each cluster independently. Note that in this case, each BS needs also to estimate statistics interference from other clusters together with intra-cluster interference.


\appendices


\end{document}